\documentclass[a4paper]{report}
\usepackage[utf8]{inputenc}
\usepackage[T1]{fontenc}
\usepackage{RJournal}
\usepackage{mathrsfs}
\usepackage{amsmath, amssymb, amsfonts, amsthm,rotating,booktabs,array}
\usepackage{bm}
\usepackage{subfigure}
\usepackage{graphicx}
\usepackage{mathabx}
\usepackage{multirow}
\usepackage{setspace}
\usepackage{mathtools}
\usepackage{multicol}
\usepackage{verbatim}
\usepackage{enumerate}

\usepackage{algorithm,algpseudocode}

\usepackage{relsize}

\DeclareMathOperator*{\argmax}{argmax}

\DeclareMathOperator{\argmin}{argmin}
\DeclareMathOperator{\tr}{tr}

\begin{document}

\sectionhead{Contributed research article}
\volume{XX}
\volnumber{YY}
\year{20ZZ}
\month{AAAA}

\begin{article}

\title{RobustCalibration: Robust Calibration of Computer Models in R}

\author{Mengyang Gu
}

\maketitle
\abstract{
Two fundamental research tasks in science and engineering are forward predictions and data inversion. This article introduces a new {R} package \CRANpkg{RobustCalibration} for Bayesian data inversion and model calibration using experiments and field observations.   Mathematical models for forward predictions are often written in computer code, and they can be computationally expensive to run. To overcome the computational bottleneck from the simulator, we implemented a statistical emulator from the \CRANpkg{RobustGaSP} package for emulating both scalar-valued or vector-valued computer model outputs. Both posterior sampling and maximum likelihood approach are implemented in the \CRANpkg{RobustCalibration} package for parameter estimation. 
 For imperfect computer models, we implement the Gaussian stochastic process and  scaled Gaussian stochastic process for modeling the discrepancy function between the reality and mathematical model. This package is applicable to various other types of field observations and models, such as repeated experiments, multiple sources of measurements and correlated measurement bias. We discuss numerical examples of calibrating mathematical models that have closed-form expressions, and differential equations solved by numerical methods.

}

\section[Introduction]{Introduction}

 Complex processes are often represented as mathematical models, {implemented} in computer code. These mathematical models are  often called \textit{computer models} or \textit{simulators}, {and are widely used to simulate different processes given a set of parameters and initial conditions} \citep{sacks1989design}. The initial conditions and model parameters may be unknown or uncertain in practice. Calibrating the computer models based on real experiments or observations is  one of the fundamental tasks in science and engineering, widely known as  \textit{data inversion} or  \textit{model calibration}. 

We follow the notation in \cite{bayarri2007computer} for defining the model calibration problem. Let $y^F(\mathbf x)$ be the real-valued field observation with a $p_x$-dimensional observable input vector $\mathbf x$. The field observation is often decomposed by $y^F(\mathbf x)=y^R(\mathbf x)+\epsilon$, where $y^R(\cdot)$ denotes a function of unknown reality, and $\epsilon$ denotes the noise, with the superscript `F' and `R' denoting the field observations and reality, respectively.  
 Scientists often model the unknown reality by a computer model, denoted by $ f^M(\mathbf x, \bm \theta)$ at the $p_x$-dimensional observable input $\mathbf  x$, and $p_{\theta}$-dimensional calibration parameters $\bm \theta $, unobservable from the experiments, with superscript `M' denoting the computer model. 
 
 When the computer model describes the reality perfectly, a field observation can be written as:
\begin{equation}
y^F(\mathbf x)= f^M(\mathbf x, \bm \theta)+ \epsilon,
\label{equ:calibration_no_discrepancy}
\end{equation}
where  $\epsilon$ is a zero-mean Gaussian noise.  We call the method by Equation (\ref{equ:calibration_no_discrepancy}) the \textit{no-discrepancy calibration}.

When  computer models are imperfect, the statistical calibration model below is often used: 
\begin{equation}
y^F(\mathbf x)= f^M(\mathbf x, \bm \theta)+ \delta(\mathbf x)+\epsilon,
\label{equ:calibration_discrepancy}
\end{equation}
where $\delta(\cdot)$ is the unobservable discrepancy between the mathematical model and reality. Equation (\ref{equ:calibration_discrepancy}) implies that the reality can be written as $y^R(\mathbf x)= f^M(\mathbf x, \bm \theta)+ \delta(\mathbf x)$ at any observable input $\mathbf x \in \mathcal X$.  The model calibration framework has been studied extensively. In \cite{kennedy2001bayesian},  the discrepancy function $\delta(\cdot)$ is modeled via a Gaussian stochastic process (GaSP), leading to more accurate  prediction based on the joint model of calibrated computer model and discrepancy compared to either using the computer model  or discrepancy model for prediction alone.  We call this method the \textit{GaSP Calibration}. The GaSP calibration has been applied in a wide range of studies  of continuous and categorical outputs \citep{bayarri2007computer,higdon2008computer,paulo2012calibration,chang2016calibrating,chang2022ice}. Both no-discrepancy calibration and GaSP calibration are implemented in the \CRANpkg{RobustCalibration} package.

The calibrated computer model output can be far from the observations in terms of $L_2$ distance when the discrepancy function is modeled by the GaSP,  \citep{arendt2012quantification}, since a large proportion of the variability in the data can be explained by discrepancy function. 
To solve this problem, we implemented the scaled Gaussian stochastic process (S-GaSP) calibration, or  \textit{S-GaSP Calibration} in the \CRANpkg{RobustCalibration} package. The S-GaSP model of discrepancy function  was first introduced in \cite{gu2018sgasp}, where more prior probability mass of the $L_2$ loss is placed on small values. Consequently, the calibrated computer model by the S-GaSP  fits the reality better in terms of $L_2$ loss than the GaSP calibration. 
Furthermore, historical data may be  used for reducing the parameter space \citep{williamson2013history}. 

 A few recent approaches aim to minimize the  $L_2$ distance between the reality and computer model \citep{tuo2015efficient,wong2017frequentist}, where the reality and discrepancy function are often estimated in two steps separately. The S-GaSP calibration bridges the two-step $L_2$ calibration with the GaSP calibration. Compared with  two-step approaches, parameters  and discrepancy $(\bm \theta, \delta(.))$ are estimated in {GaSP calibration} and {S-GaSP calibration} jointly, and the uncertainty of these estimation can be obtained based on the likelihood function of the sampling model. Furthermore, compared with the orthogonal calibration model \citep{plumlee2017bayesian}, the S-GaSP process places more prior mass on inputs leading to small values of the random $L_2$ loss, instead of more prior mass on inputs close to the critical points of the loss function. 
 
{Another concern is the computational cost of model calibration, as computer models may involve  numerical solutions of differential equations, which can be expensive to run. We utilize} a statistical emulator    for approximating computationally expensive computer models. The GaSP emulator  is a well-developed framework for emulating computationally expensive computer models and it was implemented in a few {R} packages, such as \CRANpkg{DiceKriging} \citep{roustant2012dicekriging}, \CRANpkg{GPfit} \citep{macdonald2015gpfit} and \CRANpkg{RobustGaSP}  \citep{gu2018robustgasp}. Here we specify the functions \code{rgasp}, \code{ppgasp} and \code{predict} in \CRANpkg{RobustGaSP} for emulating  computer models with scalar-valued and vector-valued outputs, respectively.

Building packages for Bayesian model calibration is much more complicated than packages of statistical emulator, as both field observations and computer models {need to be} integrated.  A few statistical packages are available for Bayesian model calibration, such as \CRANpkg{BACCO} \citep{hankin2005introducing},  \CRANpkg{SAVE} package \citep{palomo2015save}, and \CRANpkg{CaliCo} \citep{carmassi2018calico}. The \CRANpkg{CaliCo} package, for example, integrates   \CRANpkg{DiceKriging} for emulating computational expensive simulations. It offers different types of diagnostic plots based on \CRANpkg{ggplot2} \citep{wickham2011ggplot2}, and {distinct} prior choices of the parameters. In the \CRANpkg{BACCO} and \CRANpkg{SAVE}, the GaSP model of the discrepancy function \citep{kennedy2001bayesian} { is assumed, and GaSP emulators can  be used to approximate  expensive computer models}. 

Although a large number of studies were developed for Bayesian model calibration, to the authors' best knowledge, many methods implemented in \CRANpkg{RobustCalibration} have not been implemented in another software package previously. We highlight a few unique features of the \CRANpkg{RobustCalibration} package. First of all, we allow users to specify different types of output of field observations, for different scenarios through the \code{output} argument in the \code{rcalibration} function. The simplest scenario is to have a vector of fields observations. We also allow users to input a matrix  or a list of fields observations with the same or different number of replications, respectively. Efficient computation from  sufficient statistics was implemented for model calibration with replications, which can improve computation up to $k^3$ times, where $k$ is the number of repeated experiments. Second, {distinct models can be calibrated with different sets of parameters for data from multiple sources by the  \code{calibration\_MS} function}. 
Third, we implemented  emulators for approximating expensive computer models with both scalar-valued  and vector-valued outputs.  Emulators for vector-valued outputs, capable of handling a large number of temporal or spatial coordinates, were not implemented in other packages for  model calibration. Here we use the parallel partial Gaussian stochastic process  (PP-GaSP) emulator from \CRANpkg{RobustGaSP} package \citep{gu2018robustgasp} as a computationally scalable emulator of vector-valued output. The  \CRANpkg{RobustGaSP} package has been applied to various applications, such as emulating geophysical models of ground deformation   \citep{anderson2019magma}, and storm surge simulation \citep{ma2022multifidelity}. 
{Furthermore, users can choose no-discrepancy calibration,  GaSP or S-GaSP models of discrepancy functions for different scenarios. Statistical inferences by both posterior samples and the maximum likelihood estimator (MLE) are  made available for these calibration approaches by the \code{method} argument in the \code{rcalibration} function. }

The rest of the paper is organized below. {We first give an overview of the \CRANpkg{RobustCalibration} package to introduce the main functions and their usage.    We further introduce methodology and numerical examples to illustrate the implemented methods in the \CRANpkg{RobustCalibration} package.     Closed form expressions of likelihood functions, derivatives,  posterior distributions and computational algorithms are outlined in the Appendix. }

\section[An overview of \CRANpkg{RobustCalibration}]{An overview of \CRANpkg{RobustCalibration}}

\begin{figure}[t]
\centering

	\includegraphics[height=.5\textwidth,width=.98\textwidth]{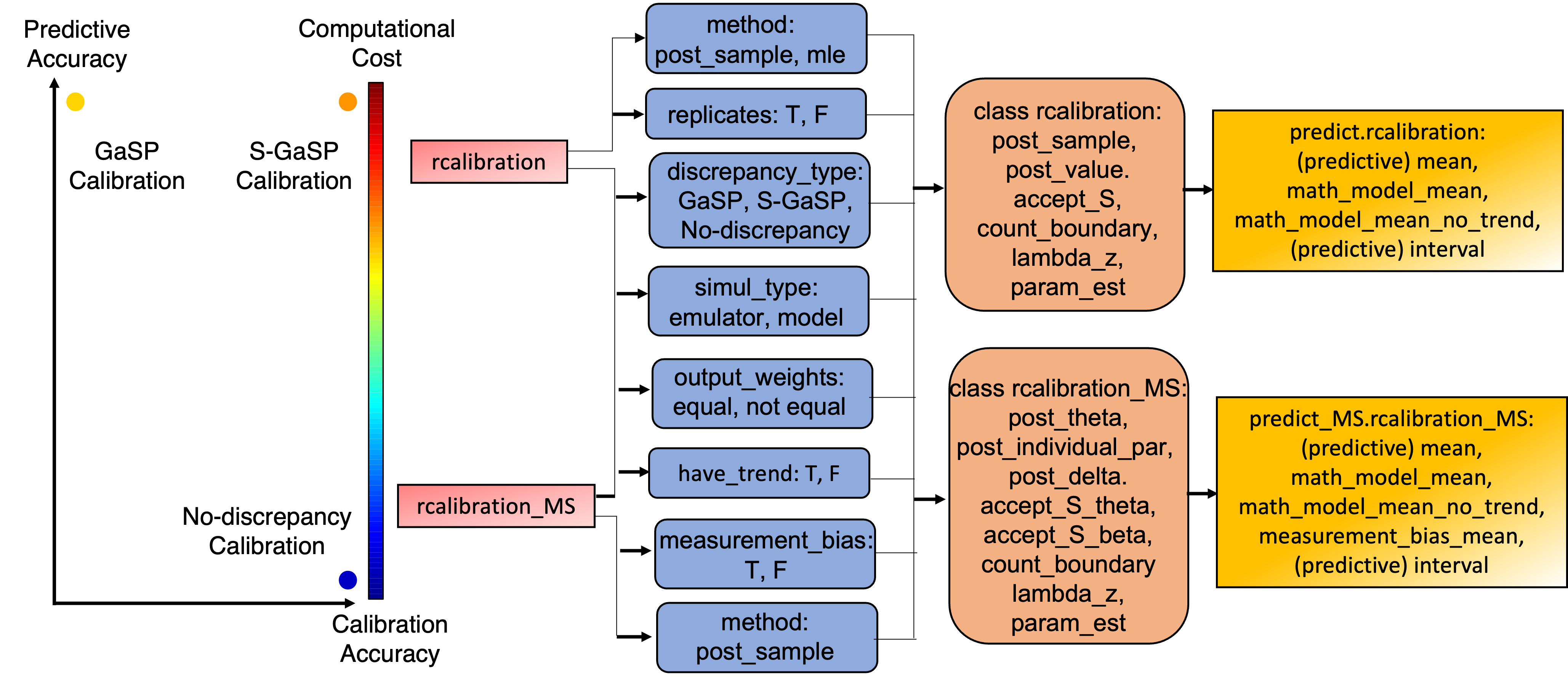}

   \caption{Schematic overview of the \CRANpkg{RobustCalibration}
 package. The left panel compares the expected predictive accuracy and calibration accuracy by different calibration methods.
 The  \code{calibration} and \code{calibration\_MS} in the right panel are two main functions for parameter estimation for observations from single and multiple sources, respectively. The arguments of these functions are given in blue boxes. The object classes  \code{calibration} and \code{calibration\_MS} (orange boxes) can be supplied in  \code{predict.rcalibration} and \code{predict\_MS.rcalibration\_MS} functions for making predictions.   
 }

 \label{fig:schematic_graph}

\end{figure}

  Figure \ref{fig:schematic_graph} gives a schematic overview of the  \CRANpkg{RobustCalibration} package. We consider predicting the reality in two ways:  using both calibrated computer model and discrepancy (predictive accuracy), and the calibrated computer model alone (calibration accuracy), both evaluated in terms of the $L_2$ loss. The left panel gives comparison of accuracy and computational cost between the implemented methods. {
  The no-discrepancy calibration is  faster than GaSP and S-GaSP calibration, as one does not need to compute the inversion and log determinant of the covariance matrix of the field data, unlike GaSP calibration and S-GaSP calibration.} 
    Because model discrepancy is modeled,  predictive accuracy by GaSP and S-GaSP calibration is typically higher than no-discrepancy calibration. The \CRANpkg{RobustCalibration} package can handle model calibration and prediction at a wide range of settings. We first introduce the main functions of the package.

\subsection{Main functions}
The  \CRANpkg{RobustCalibration} package can be used for estimating unobservable parameters from computer models and  predicting reality. The conventional model calibration and prediction from a single source of field observations are achieved by \code{rcalibration} and \code{predict.rcalibration}, as shown in upper blue and  yellow boxes in Figure \ref{fig:schematic_graph}, respectively. The function \code{rcalibration} allows users to call {either} a Markov chain Monte Carlo (MCMC) algorithm for posterior sampling or a numerical optimization algorithm for computing the maximum likelihood estimator (MLE)  of the parameters. Additional arguments can be specified to handle observations from repeated experiments, select a different trend or discrepancy function, and build a surrogate model for approximating  expensive computer models, shown in  Figure \ref{fig:schematic_graph}.
 The \code{rcalibration} function returns an object
of the \code{rcalibration} S4 class with posterior samples or MLE of the parameters, {shown in the upper orange box in Figure \ref{fig:schematic_graph}}. 
Then \code{rcalibration} S4 class is used as the input in the \code{predict.rcalibration} function to perform predictions on a set of test inputs. The  \code{predict.rcalibration} function returns  a \code{predictobj.rcalibration} S4 class, which contains the predictive mean of the calibrated computer model,  estimated trend, and  discrepancy function. The estimated interval at a given quantile can also be returned by specifying the vector of the \code{interval\_est} argument in the \code{predict.raclibration} function.  
 
  In some applications, we have different sources or types of observations. For instance, in calibrating geophysical models, time series of continuous GPS observations and satellite radar interferogram can be jointly used  to estimate unobservable parameters \citep{anderson2019magma}. 
  In the \CRANpkg{RobustCalibration} package, we implement parameter estimation for observations from multiple sources by the \code{rcalibration\_MS} function, which returns a  \code{rcalibration\_MS} S4 class that contains posterior samples, shown in Figure \ref{fig:schematic_graph}. {Since each source of observations can induce a separate set of model parameters, making the dimension of the parameter space large, the MLE is unstable, and as such is not implemented in \code{rcalibration\_MS}.} The \code{rcalibration\_MS} class can be used as an input into the \code{predict\_MS.rcalibration\_MS} function for predicting the reality. 
  
  \subsection{The \code{rcalibration} function} 
To use the \code{rcalibration}  function, users need to specify 4 inputs: 1) \code{design}, an $n\times p_x$ matrix of observable inputs, 2) \code{observations}, field observations that can have three types specified below,  3) \code{theta\_range}, a $p_{\theta}\times 2$ matrix of the lower and upper bounds of calibration parameters, and 4) either \code{math\_model}, a function of mathematical model, or \code{input\_simul} and \code{output\_simul}, simulation runs by for building the emulator. The \CRANpkg{RobustCalibration} package can handle three different kinds of field observations as specified by the argument \code{observations}. First, one can input an n-\code{vector} $(y^F(\mathbf x_1),...,y^F(\mathbf x_n))^T$, when one field datum is available at each observable input. Second, when $k$ repeated measurements are available for each observable input, one can input an $n\times k$ \code{matrix} of field observations, where each row contains $k$ replicates of one observable input. Third, one can input a \code{list}, where each element contains  $k_i$ replications at the $i$th observable input, for $i=1,...,n$.

An important feature of the \CRANpkg{RobustCalibration} package is the flexible specification of the computer model. Users can  specify a  mathematical model by supplying a generic \code{R} function through the argument \code{math\_model} and  selecting the default choice of the simulator \code{simul\_type=1}. The second choice is to use an emulator to predict the computer model by choosing \code{simul\_type=0}. Users need to input a  set of simulation runs by \code{input\_simul} and \code{output\_simul} to call an emulator by the \CRANpkg{RobustGaSP} package for this choice.  {Users can either give  a $D\times (p_x+p_{\theta})$ matrix of input and a $D$-vector of output  to \code{input\_simul} and \code{output\_simul} arguments, for calling \code{rgasp} function from \CRANpkg{RobustGaSP} for emulation,  or give  a  $D \times p_{\theta} $ matrix of input and a $D\times p_x $ matrix of output to \code{input\_simul} and \code{output\_simul}, to call for \code{ppgasp} function from  \CRANpkg{RobustGaSP} for emulation}. To account for stochastic error or numerical approximation error in the simulator, we also allow users to specify a nugget parameter estimated by the simulator data by the argument \code{simul\_nug=T}.

The \code{rcalibration} function contains a few  optional arguments. First, users can specify GaSP or S-GaSP  discrepancy models  by the argument \code{discrepancy\_type=`GaSP'}  or \code{discrepancy\_type=`S-GaSP'}, respectively. The default choice is to assume a S-GaSP discrepancy model. The model without a discrepancy can be specified by the argument \code{discrepancy\_type=`no-discrepancy'}. Second, a trend can be specified by the argument \code{X}. By default, the trend (or mean) of the calibration model is zero. Third, the  parameter estimation approach can be specified by the \code{method} argument. The default choice is \code{method=`post\_sample'}, where the posterior samples will be drawn, and stored in the slot \code{post\_sample} of the \code{rcalibration} class.    The MLE can be specified by the argument \code{method=`mle'}, and the estimated parameters are   stored in the slot \code{param\_est} of the \code{rcalibration} class.  Users can specify the number of MCMC samples and burn-in samples by the argument \code{S} and \code{S\_0}, respectively. A vector containing the standard deviation of the proposal distribution for the calibration parameters, the logarithm of the inverse range parameters, and the logarithm of the nugget parameter can be specified by the \code{sd\_proposal} argument. Furthermore, users can ``thin'' the MCMC samples and only record a subset by  the argument \code{thinning}. For instance, \code{thinning=5} means only 1/5 of the posterior samples will be recorded.  
Besides, the inverse variances of noises in the field observations can be specified by the \code{output\_weights}. Finally,   arguments \code{initial\_values} and \code{num\_initial\_starts} can be specified in numerical optimization to find MLE, and when posterior sampling is used,   initial values of the calibration parameters can be specified through the argument \code{initial\_values}.

The object \code{rcalibration} created by the  \code{rcalibration} function have a few key properties. First of all, if we have \code{method='post\_sample'}, the after burn-in posterior samples of parameters will be stored in the slot \code{post\_sample}, where each row contains posterior samples in one iteration. The first $p_\theta$ columns contain posterior samples of the calibration parameters $\bm \theta$. In the no-discrepancy calibration, the $p_\theta+1$ column of \code{post\_sample} contains posterior draws of the noise variance $\sigma^2_0$ and the $p_\theta+2$ to $p_\theta+q+1$ columns contain the trend parameters $\bm \theta_m$, for a non-zero mean basis. In the GaSP or S-GaSP calibration, the $p_\theta+1$ to $p_\theta+p_x+1$ columns record  posterior samples of the log inverse range and log nugget parameters. The $p_\theta+p_x+2$ to $p_\theta+p_x+q$ columns record the noise variance parameter and trend parameters if they are specified. 
The parameter $\lambda_z$ in S-GaSP are recorded in slot \code{lambda\_z}. 
 The indices of the  accepted proposed samples are recorded in slots \code{accept\_S} as one of the diagnostic statistics. 

  \subsection{The \code{predict.rcalibration} function} 
  
  After calling the \code{rcalibration} function, an object of the \code{rcalibration} S4 class will be created, and it can be used as an input into the \code{predict.rcalibration} function for predicting the reality on a specified matrix of test input using the argument \code{testing\_input}.     Besides, {if the mean basis is specified by the argument  \code{X}   in \code{rcalibration} function}, the user also needs to specify the mean basis of the reality at the test inputs by the argument \code{X\_testing} in function \code{predict.rcalibration}. If the emulator was not constructed in  \code{rcalibration}, users also need to specify the mathematical model by the argument \code{math\_model} in the \code{predict.rcalibration} function.
   Finally, a predictive interval can be specified by the argument \code{interval\_est}. For instance, the 95\% predictive interval can be obtained by the  \code{interval\_est=c(0.025,0.975)} argument. The interval of the field data will be computed  if \code{interval\_data=T}. Otherwise, the predictive interval of the reality will be computed. 
 If the {variance} parameter of the noise in field data was specified by the \code{output\_weights}, users {should} also specify the variance of test data by \code{testing\_output\_weights}, which affects the predictive interval of test data.  
  
  After calling  \code{predict.rcalibration} function, an object \code{predictobj.rcalibration}  will be created and it  contains three different predictors for reality. The slot \code{math\_model\_mean\_no\_trend} gives the predictive mean based on the calibrated computer model ($f^M$). The slot \code{math\_model\_mean}  gives the predictive mean based on the calibrated computer model and the estimated trend ($f^M+\mu$), if the basis functions of the trend at the observable input and the test input are specified through \code{X} and \code{X\_testing} in \code{rcalibration} and \code{rcalibration.predict}, respectively. The slot \code{mean} gives the predictive mean based on the calibrated computer model, estimated trend, and discrepancy  ($f^M+\mu+\delta$). If \code{interval\_est} is specified, a matrix of the intervals will be created for quantifying the uncertainty of predictions. 
  
  \subsection{The \code{rcalibration\_MS} function and the \code{predict\_MS.rcalibration\_MS} function}

Model calibration and predictions using multiple sources or different types of data are implemented by  \code{rcalibration\_MS} and \code{rcalibration\_MS.predict\_MS} functions, respectively. One needs to specify 4 inputs: \code{design}, \code{observations}, \code{theta\_range} and a form of mathematical model. Suppose we have a data set produced by $k$ sources of observations.  The \code{design} is a list of $k$ elements, where each element is a matrix of observable input for each source. The  argument \code{observations} takes a list of field observations, where each element is a vector of field observations for each source. In principle, one can have different number of observations from each source and the type of the observations may not be the same. 
The  argument \code{theta\_range} is a $p_{\theta}\times 2$ matrix of  range of parameters, where {the $i$th row} contains the minimum and maximum values of the $i$th coordinate of the calibration parameter vector $\bm \theta$. Furthermore, if closed form expressions of mathematical models are available for each source of data, users can input a list of functions into the \code{math\_model} for each source of data. We also allow users to emulate the expensive simulation. In this scenario, one needs to let \code{simul\_type} be a vector of $1$s, indicating emulators will be called for each computer model. 

 Additional arguments can be specified in \code{rcalibration\_MS}. For instance, the argument \code{index\_theta} takes a list of indices for the associated calibration parameter for each computer model. Suppose we have three calibration parameters $\bm \theta=(\theta_1,\theta_2,\theta_3)$, and the computer model for first source of data is related to the first two calibration parameters and  the computer model for second source of data is related to second two calibration parameters. Then \code{index\_theta} is a list where \code{index\_theta[[1]]=c(1,2)}, and \code{index\_theta[[2]]=c(2,3)}. More arguments will be introduced along with the examples. 
   After calling \code{rcalibration\_MS}, an S4 class \code{rcalibration\_MS} will be built, and used as an input to \code{predict\_MS.rcalibration\_MS} for making predictions.

\section[Methods and examples]{Methods and examples}

\subsection{No-discrepancy calibration}
{Let us start with the simplest method: no-discrepancy calibration}. First, suppose we have a vector of real-valued field observations $\mathbf y^F=(y^F(\mathbf x_1),...,y^F(\mathbf x_n))^T$ at $n$ observable inputs, and the corresponding computer model output is denoted by $\mathbf f^M_{\bm \theta}=(f^M(\mathbf x_1,\bm \theta),...,f^M(\mathbf x_n,\bm \theta))^T$ for any calibration parameters $\bm \theta$. The vector of measurement noise is assumed to follow $\bm \epsilon \sim \mathcal{MN}(0,\, \sigma^2_0 \bm \Lambda)$, where the covariance of the noise is diagonal with the $i$th term being $\sigma^2_0\Lambda_{ii}$ and $ \mathcal{MN}$ denote the multivariate normal distribution. 
 In the default setting, $\sigma^2_0$ is estimated from the data and $ \bm \Lambda=\mathbf I_n$, where $\mathbf I_n$ is an $n\times n$ identity matrix. Users can manually specify the inverse of the diagonal terms of $\bm \Lambda$ by the \code{output\_weights} argument in \code{rcalibration} and \code{rcalibration\_MS} functions, as the variances of measurement errors can be different. We assume the diagonal matrix $\bm \Lambda$ is given in this section.

  For a no-discrepancy calibration model in (\ref{equ:calibration_no_discrepancy}),  the MLE of the variance of the noise is $\hat \sigma^2_0=S^2_0/n$ with $S^2_0=(\mathbf y^F-\mathbf f^M_{\theta} )^T \bm \Lambda^{-1}(\mathbf y^F-\mathbf f^M_{\theta} )$.  The likelihood function is given in the Appendix. As $\bm \Lambda$ is a diagonal matrix with diagonal terms denoted as $\Lambda_{ii}:=1/w_i$, the profile likelihood  follows
\[ \mathcal L( \bm \theta\mid  \hat \sigma^2_0 ) \propto  \left\{ \sum^n_{i=1} w_i \left(y^F(\mathbf x_i)- f^M(\mathbf x_i, \bm \theta)\right)^2\right\}^{-n/2}, \]
where $w_i=1$ is used in the default scenario.   Maximizing the profile likelihood of a no-discrepancy model is equivalent to minimizing the weighted squared error  $\sum^n_{i=1} w_i \left(y^F(\mathbf x_i)- f^M(\mathbf x_i, \bm \theta)\right)^2$. We numerically find $\bm \theta$ by the low-storage quasi-Newton optimization method \citep{liu1989limited} implemented in \code{lbfgs} function in the \CRANpkg{nloptr} package (\cite{nloptr2014}) for optimization. 
  
  In the \CRANpkg{RobustCalibration} package, we allow the users to specify the trend or mean function modeled as $\bm \mu(\mathbf x_i)=  \mathbf h(\mathbf x) \bm \theta_{m}=\sum^q_{t=1}h_t(\mathbf x)  \theta_{m,t} $ for any $\mathbf x$, where $\bm \theta_{m}=(\theta_{m,1},...,\theta_{m,q})^T$ is a vector of trend parameters. Maximizing the profile likelihood of a no-discrepancy model with a trend is equivalent to minimizing the squared error loss: 
   \[ \bm {\hat \theta}^{LS}_{m}=\argmin_{\bm \theta}\sum^n_{i=1} w_i \left(y^F(\mathbf x_i)- f^M(\mathbf x_i, \bm \theta) - \mathbf h(\mathbf x_i)\hat {\bm \theta}^{LS}_m    \right)^2, \] 
where the solution follows $\hat {\bm \theta}^{LS}_{m}= (\mathbf H^T \bm \Lambda^{-1} \mathbf H)^{-1}\mathbf H^T \bm \Lambda^{-1}( \mathbf y^F-\mathbf f^M_{\theta})$ with   $\mathbf H=(\mathbf h^T(\mathbf x_1),..., \mathbf h^T(\mathbf x_n))^T$ being an $n\times q$ matrix of the mean basis.     
   
After calling the  \code{rcalibration} function with \code{method='mle'}, the point estimator $\bm {\hat \theta}_{LS}$ is recorded in the slot \code{param\_est} in the \code{rcalibration} class.  The MLE is  a computationally cheap way to obtain estimates of calibration parameter vector $\bm \theta$.  After obtaining the MLE, we can plug the MLE into the computer model for predicting reality: $f^M(\mathbf x_i,  \bm {\hat \theta}) + h(\mathbf x_i)\hat {\bm \theta}^{LS}_m$, {using the \code{predict} function}.

We also implement the MCMC algorithm for Bayesian inference.
The  posterior distribution follows 
\begin{equation}
p(\bm \theta_m, \sigma^2_0, \bm \theta\mid \mathbf y^F) \propto  p(\mathbf y^F \mid \bm \theta, {\bm  \theta_m},  \sigma^2_0 ) \pi(\bm \theta) \pi({\bm  \theta_m},  \sigma^2_0).
\label{equ:post}
\end{equation} 
We assume an objective prior for the mean and variance parameters $\pi(\bm \theta_m, \sigma^2_0)\propto 1/\sigma^2_0$.  The posterior distribution $p(\bm \theta_m, \sigma^2_0 \mid \bm \theta, \mathbf y^F)$ can be sampled by the Gibbs algorithm, since the prior is conjugate. We assume that the default choice of the prior of the calibration parameters $\pi(\bm \theta)$ is uniform over the parameter space and sample $(\bm \theta \mid \mathbf y^F,\bm \theta_m, \sigma^2_0)$  using the Metropolis algorithm. The posterior distributions and predictions from the posterior samples will be discussed in the Appendix.

 Here we show one example from \cite{bayarri2007framework}, where data are observed from unknown reality:  $y^F(x)=3.5\exp(-1.7x)+1.5+\epsilon$, where $\epsilon \sim \mathcal N(0,0.3^2)$. We have 30 observations at 10 inputs in the domain $x_i \in [0,3]$, for $i=1,2,...,10$, each containing $3$ replications. The 
computer model is $f^M(x,\theta)= 5 \exp(-\theta x)$ with $\theta \in [0,\, 50]$.  The goal is to estimate the calibration parameters and predict the unknown reality ($3.5\exp(-1.7x)+1.5$) at $x\in [0,5]$. 
The observable inputs and observations from  \cite{bayarri2007framework} can be generated by the code below. 
\begin{example}
R> input=c(.110, .432, .754, 1.077, 1.399, 1.721, 2.043, 2.366, 2.688,3.010)
R> n=length(input)
R> k=3
R> output=t(matrix(c(4.730,4.720,4.234,3.177,2.966,3.653,1.970,2.267,2.084,2.079,
+                2.409,2.371,  1.908,1.665,1.685, 1.773,1.603,1.922,1.370,1.661,
+                1.757, 1.868,1.505,1.638,1.390, 1.275,1.679,1.461,1.157,1.530),k,n))
R> Bayarri_07<-function(x,theta){
+   5*exp(-x*theta) 
+ }
R> theta_range=matrix(c(0,50),1,2) 
\end{example}
Here the mathematical model has a closed-form expression so we code it as a function called \code{Bayarri\_07}. The parameter range of calibration parameter $\theta$ is given by a vector called \code{theta\_range}. 

 The no-discrepancy calibration is implemented by the code below. 
\begin{example}
R> set.seed(1)
R> X=matrix(1,n,1)
R> m_no_discrepancy=rcalibration(input, output,math_model = Bayarri_07,
+                                theta_range = theta_range, X =X, 
+                                have_trend = T,discrepancy_type = 'no-discrepancy')
R> m_no_discrepancy
type of discrepancy function:  no-discrepancy 
number of after burn-in posterior samples:  8000 
0.1385 of proposed calibration parameters are accepted 
median and 95\% posterior credible interval of calibration parameter  1 are 
2.951997 2.264356 3.930725 
\end{example}
Here we {use} a constant mean basis using the mean (or trend) argument \code{X}, such that the mathematical model is $f^M(x,\theta)+\mu$. We draw $10,000$ posterior samples with the first $2,000$ samples used as burn-in samples. One can adjust the number of posterior and burn-in samples by argument \code{S} and \code{S\_0} in the \code{calibration} function, respectively. We found that the median of the posterior samples of $\theta$ is $2.95$, and the $95\%$ posterior interval is around $(2.26, 3.93)$. Around $14\%$ of the samples were accepted by the Metropolis algorithm. Here the first $p_{\theta}$ terms in \code{sd\_proposal} is the standard deviation of the proposal distribution of the calibration parameter, where the default choice is $0.05$, and the next $p_x+1$ parameters are the standard deviation of the inverse logarithm of the range parameter and nugget parameters in GaSP and S-GaSP calibration, discussed in the next subsection. {The code below reduces \code{sd\_proposal} to  $0.025$ of the range of the parameter space to make a larger proportion of the posterior samples accepted by the algorithm:}
  \begin{example}
R> m_no_discrepancy_small_sd=rcalibration(input, output,math_model = Bayarri_07,
+                           theta_range = theta_range,
+                           sd_proposal = c(rep(0.025,dim(theta_range)[1]),
+                                            rep(0.25,dim(as.matrix(input) )[2]),0.25),
+                           X =X, have_trend = T,discrepancy_type = 'no-discrepancy')
R> m_no_discrepancy_small_sd
type of discrepancy function:  no-discrepancy 
number of after burn-in posterior samples:  8000 
0.2613 of proposed calibration parameters are accepted 
median and 95\% posterior credible interval of calibration parameter  1 are 
2.935007 2.19352 3.933492 
\end{example}
Around 26\% of the posterior samples are accepted.  In this example, the median and $95\%$ posterior credible intervals are similar to the one having a proposal distribution with smaller standard deviation.

{The following code gives the predictions and the $95\%$ predictive interval of the reality for 200 test inputs equally spaced at $x\in [0,5]$. Since we use a constant trend in model calibration, we also specify a constant trend for making predictions, through the argument \code{X\_testing} in the \code{predict} function: }
\begin{example}
R> testing_input=seq(0,5,5/199)
R> X_testing=matrix(1,length(testing_input),1)
R> m_no_discrepancy_pred=predict(m_no_discrepancy,testing_input,math_model=Bayarri_07,
+                               interval_est=c(0.025, 0.975),X_testing=X_testing) 
\end{example}

\begin{figure}[t]
\centering
  \begin{tabular}{ccc}

	\hspace{-.2in} \includegraphics[scale=.7]{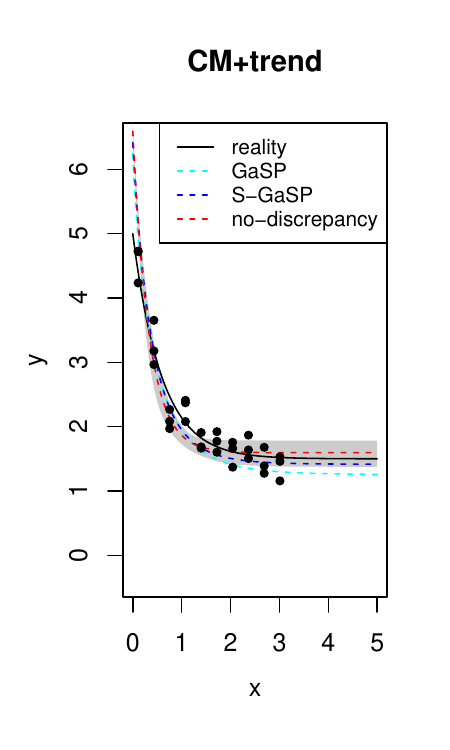}
	\hspace{-.3in} \includegraphics[scale=.7]{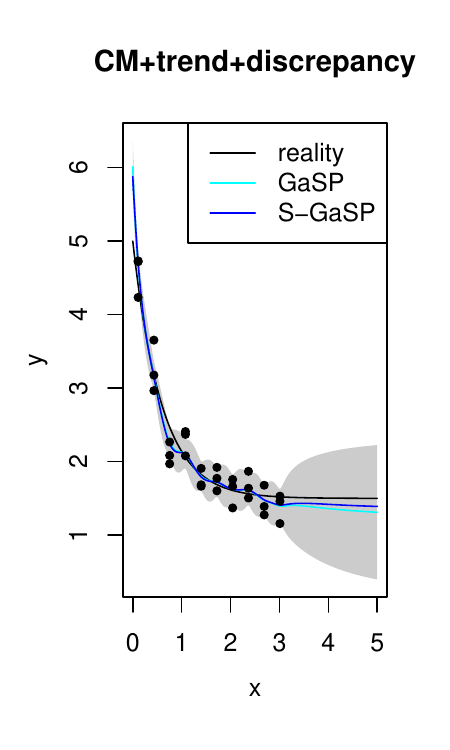}
		\hspace{-.3in} 	\includegraphics[scale=.7]{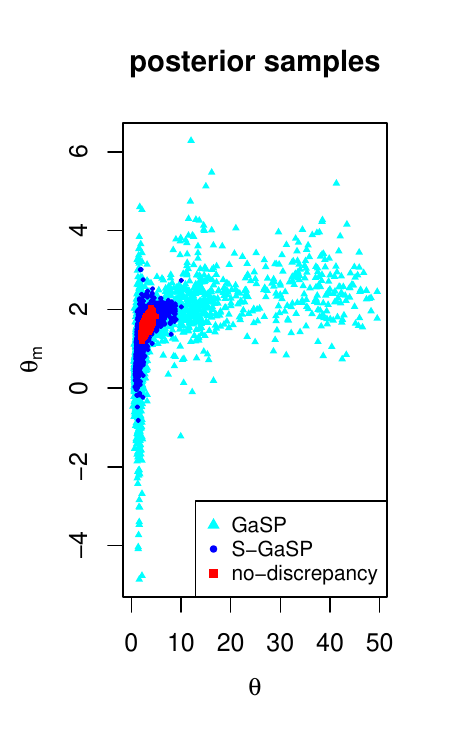}
\end{tabular}
   \caption{Predictions and posterior samples by the \CRANpkg{RobustCalibration} package for the example introduced in \cite{bayarri2007framework}. In the left and middle panels, the unknown reality is graphed as black curves, and field observations that contain with 3 replicates at each observable input are plotted as black {dots}.  In the left panel, predictions of reality based on calibrated computer model with trend by GaSP, S-GaSP and no-discrepancy calibration are graphed as light blue, blue and red curves, respectively. `CM'  denotes the computer model and the shaded area is the $95\%$ posterior predictive interval by the no-discrepancy calibration.  In the middle panel, predictions based on the calibrated computer model, discrepancy, and the trend by GaSP and S-GaSP are graphed as light blue and blue curves, respectively. The shaded area is the  $95\%$  predictive interval by the GaSP calibration.  After burn-in posterior samples of calibration parameter $\theta$ and mean parameter $\theta_m$  are plotted in the right panel.  }

 \label{fig:bayarri_2007}
\end{figure}

 Posterior samples of the calibration parameter $\theta$ and the mean parameter $\theta_m$ in the no-discrepancy calibration are plotted  as the red rectangles in the right panel in Figure \ref{fig:bayarri_2007}. The predictive mean and 95\% predictive interval  are graphed as the red curve and grey shaded area in the left panel,  respectively. Compared to models with an estimated discrepancy function, the posterior samples from the no-discrepancy calibration are more concentrated, and the average length of 95\% predictive interval is shorter. However, a large proportion of reality at $x\in[0, 1.5]$ is not covered by the 95\% posterior credible interval, due to large discrepancy between the mathematical model  and reality  in this  domain. 
Next, we introduce two discrepancy models to solve this problem.  

\subsection{Gaussian stochastic process models of discrepancy functions}

 The discrepancy function can be  modeled as a GaSP, meaning that for any $\{\mathbf x_1,...,\mathbf x_n\}$, the marginal distribution of the discrepancy function follows a multivariate normal distribution: 
\[(\delta(\mathbf x_1),..., \delta(\mathbf x_n))^T \mid  \bm \mu, \sigma^2, \mathbf R \sim  \mathcal{MN}(\bm \mu,\,  \sigma^2 \mathbf R), \]
where $\bm \mu=(\mu(\mathbf x_1),...,\mu(\mathbf x_n))^T $ is a vector of the mean (or trend),  $\sigma^2$ is a variance parameter, and $\mathbf R$ is the correlation matrix between outputs. 

Similar to the no-discrepancy calibration, we allow user to model the trend as $\bm \mu(\mathbf x_i)=  \mathbf h(\mathbf x_i) \bm \theta_{m}=\sum^q_{t=1}h_t(\mathbf x_i)  \theta_{m,t} $, where  $\mathbf h(\mathbf x_i)$ is a row vector of mean basis and $\bm \theta_{m}$ is a vector of trend parameters. Since the trend is often modeled in the computer model, the default value of the trend  parameters is $\mathbf 0$. 

The $(i,j)$th element of the correlation matrix is modeled by a kernel function $R_{i,j}=K(\mathbf x_i, \mathbf x_j)$. We implement the separable kernel  $K(\mathbf x_i, \mathbf x_j)=\prod^{p_x}_{l=1}K_l(\mathbf x_i, \mathbf x_j)$, where $K_l$ models the correlation between the $l$th coordinate of the observable input, for $l=1,...,p_x$. Some frequently used kernel functions are listed in Table \ref{tab:kernel}. The range parameters $\bm \gamma$ are estimated by default.

 \begin{table}[t]
\begin{center}
\begin{tabular}{ll}
  \hline
  GaSP calibration & Product kernel  ($K(\mathbf d)=\prod^{p_x}_{l=1}K_l(d_l)$) \\

  \hline
  Mat{\'e}rn $\alpha=5/2$                 & $K_l(d_l)=\left(1+\frac{\sqrt{5}d_l}{\gamma_l}+\frac{5d^2_l}{3\gamma_l^2}\right)\exp\left(-\frac{\sqrt{5}d_l}{\gamma_l}\right)$ \\
  Mat{\'e}rn $\alpha=3/2$                 & $K_l(d_l)=\left(1+\frac{\sqrt{3}d_l}{\gamma_l}\right)\exp\left(-\frac{\sqrt{3}d_l}{\gamma_l}\right)$ \\
  Power exponential                  &  $K_l(d_l)=\exp\left\{-\left(\frac{d_l}{\gamma_l}\right)^{\alpha_l}\right\}$, $0<\alpha_l\leq 2$ \\
    \hline
    S-GaSP calibration   &Scaled kernel ($K_{Z_d}(\mathbf x_a,\mathbf x_b )=K(\mathbf x_a,\mathbf x_b ) - \mathbf r^T(\mathbf x_a)  {\mathbf R}_z^{-1} \mathbf r(\mathbf x_b)$) \\
    \hline
\end{tabular}
\end{center}
   \caption{Kernel functions  implemented in \CRANpkg{RobustCalibration}. For any $\mathbf x_a, \mathbf x_b \in \mathcal X$, denote $\mathbf d=\mathbf x_a-\mathbf x_d=(d_1,...,d_{p_x})^T$. For any kernel $K$, the discretized scaled kernel $K_{Z_d}(\mathbf x_a,\mathbf x_b )$ with discretization points on observed points $\mathbf x_1,..., \mathbf x_n$ is implemented in  the S-GaSP calibration, where $  {\mathbf R}_z:=\mathbf R+n\mathbf I_n / \lambda_z $, and $\mathbf r(\mathbf x)=(K(\mathbf x, \mathbf x_1),...,K(\mathbf x, \mathbf x_n))^T$ for any $\mathbf x$.     }
   \label{tab:kernel}
\end{table}

 We denote the nugget parameter by  the noise variance to signal variance ratio $\eta=\sigma^2_0/\sigma^2$.   Marginalizing out the random discrepancy function $\delta(\cdot)$, one has $\mathbf y^F \sim \mathcal{MN}(\mathbf f^M_{\theta}+\mathbf H \bm \theta_{m},\, \sigma^2_0 \tilde {\mathbf R})$, where $\tilde {\mathbf R}=\mathbf R/\eta+ \bm \Lambda$. Here by default $\bm \Lambda=\mathbf I_n$, and we allow users to specify the inverse diagonal terms of $\bm \Lambda$ by the argument \code{weights}: $w_i=1/\Lambda_{ii}$, for $i=1,...,n$.

The MLE of the trend and noise variance parameters in the GaSP calibration follows $\hat {\bm \theta}_m= \left( \mathbf H^T {{\mathbf { \tilde R}}}^{-1}  \mathbf H \right)^{-1} \mathbf H^T {{\mathbf { \tilde R}}}^{-1}{(\mathbf{y}-\mathbf f^M_{\theta})}$ and $ \hat{\sigma}^2_0 =S^2_{K}/n$, where $S^2_K=  (\mathbf{y}-\mathbf f^M_{\theta}- \mathbf H^T \hat{\bm \theta}_m  )^T  {{\mathbf { \tilde R}}}^{-1}(\mathbf{y}-\mathbf f^M_{\theta}- \mathbf H^T \hat{\bm \theta}_m  )$.  Plugging in the MLE of  trend and noise variance parameters, one can numerically maximize the profile likelihood to obtain the calibration, kernel and nugget parameters in the GaSP calibration: 
\begin{equation}
(\hat {\bm \theta}, \hat {\bm \gamma}, \hat \eta ) =\argmax_{(\bm \theta, \bm \gamma, \eta)} \mathcal L_K( \bm \theta, \bm \gamma, \eta \mid \hat {\bm \theta}_m, \hat \sigma^2_0 )  
\label{equ:MLE_GaSP} 
\end{equation}
where the closed form expression of the profile likelihood $\mathcal L_K( \bm \theta, \bm \gamma, \eta \mid { \hat {\bm \theta}_m}, \hat \sigma^2_0 )  $ with kernel $K(.,.)$  is  given in the Appendix.

 After obtaining the estimation of the parameters, the predictive distribution of reality on any $\mathbf x^*$ can be obtained by plugging in the estimator 
\begin{equation}
 y^R(\mathbf x^*)\mid \mathbf y^F, \hat{\bm \theta}_m, \hat \sigma^2_0, \hat {\bm \theta}, \hat {\bm \gamma}, \hat \eta \sim \mathcal N( \hat \mu(\mathbf x^*), \hat \sigma^2_0(K^*/\hat \eta+   \Lambda^*) ), 
 \label{equ:predict_dist_gasp}
 \end{equation}
where $ \Lambda^*$ is the weight at $\mathbf x^*$ set to be 1 by default, and  
\begin{align*}
\hat \mu(\mathbf x^*)&=f^M(\mathbf x^*, \hat {\bm \theta} )+\mathbf h(\mathbf x^*) \hat{\bm \theta}_m+ \mathbf r^T(\mathbf x^*)  \tilde {\mathbf R}^{-1}(\mathbf y^F- \mathbf f(\mathbf x_{1:n},\hat {\bm \theta} )-\mathbf H \hat{\bm \theta}_m ), \\
K^*&=K(\mathbf x^*, \mathbf x^*)- \mathbf r^T(\mathbf x^*)  \tilde {\mathbf R}^{-1} \mathbf r(\mathbf x^*), 
  \end{align*}
with  $\mathbf r(\mathbf x^*)=(K(\mathbf x^*, \mathbf x_1),...,K(\mathbf x^*, \mathbf x_n))^T$.  {Though the predictive mean $\hat \mu(\mathbf x^*)$ may be the most accurate for predictions, the  calibrated computer model $f^M(\mathbf x^*, \hat {\bm \theta} )$   is sometimes of interest for predicting the reality, due to its interpretability}. Thus, we recorded three slots for predicting the reality: 1) the calibrated computer model $f^M(\mathbf x^*, \hat {\bm \theta} )$ by \code{math\_model\_mean\_no\_trend}, 2) The calibrated computer model with trend $f^M(\mathbf x^*, \hat {\bm \theta} )+\mathbf h(\mathbf x^*) \hat {\bm \theta}_m$  by \code{math\_model\_mean} , and 3) the predictive mean $\hat \mu(\mathbf x^*)$ by \code{mean}. Furthermore, users can also output the predictive interval based on  (\ref{equ:predict_dist_gasp}) by  specifying  \code{interval\_est} in the \code{predict.rcalibration} function. For example, setting   \code{interval\_est=c(0.025,0.975)}  gives the 95\% predictive interval of the reality. Furthermore, if \code{interval\_data=T}, the predictive interval of the data will be computed. 

Numerically computing the MLE  can be unstable  as the profile likelihood  $\mathcal L_K( \bm \theta, \bm \gamma, \eta\mid {\bm \hat \theta_m}, \hat \sigma^2_0 )$ is {typically} nonlinear and nonconvex with respect to the parameters $(\bm \theta, \bm \gamma, \eta)$. 
To obtain the uncertainty assessment of the estimates and to avoid instability in the numerical search in computing the MLE, we implement the MCMC algorithm for Bayesian inference. The posterior distribution is  
\begin{equation}
p(\bm \theta_m, \sigma^2_0, \bm \theta, \bm \gamma, \eta\mid \mathbf y^F) \propto  p_K(\mathbf y^F \mid \bm \theta, \bm \gamma, \eta, {\bm  \theta_m},  \sigma^2_0 ) \pi(\bm \theta) \pi({\bm  \theta_m},  \sigma^2_0, \bm \gamma,\eta), 
\label{equ:post}
\end{equation}
where $p_K(.)$ is a multivariate normal density with covariance function of the discrepancy being $\sigma^2K(\cdot,\cdot)$, and $\pi(\bm \theta)$ is the prior of the calibration parameters, assumed to be uniform over the parameter space. The prior of the mean, variance, kernel and nugget parameters takes the form below  
\[  \pi({\bm  \theta_m},  \sigma^2_0, \bm \gamma,\eta) \propto \frac{\pi(\bm \gamma,\eta)}{\sigma^2_0},   \]
where the usual reference prior of the trend and variance parameters $\pi({\bm  \theta_m},  \sigma^2_0)\propto 1/\sigma^2_0$ is assumed. We use jointly robust prior  for the kernel and nugget parameters $\pi(\bm \gamma,\eta)$,  as it has a similar tail density decay rate as  the reference prior, but it is easier and more robust to compute \citep{gu2019jointly}.  Closed form expressions of the posterior distributions and MCMC algorithm are provided in the Appendix.

In \CRANpkg{RobustCalibration}, after burn-in posterior samples $(\bm \theta^{(i)}_m, (\sigma^2_0)^{(i)}, \bm \theta^{(i)}, \bm \gamma^{(i)}, \eta^{(i)})$ are recorded for $i=S_0+1,...,S$. Users can  record a subset of posterior samples {to reduce storage} by thinning the Markov chain. For instance,  \code{thinning=5} in the \code{rcalibration} function  {records $1/5$ of  posterior samples}.

  We also output three types of predictions.    For instance, suppose we obtain $S$ posterior samples with first $S_0$ samples used as burn-in samples, the predictive mean of the reality can be computed by 
\[ \hat \mu(\mathbf x^*)=\sum^S_{i=S_0+1} f^M(\mathbf x^*,  {\bm \theta}^{(i)} )+\mathbf h(\mathbf x^*)  {\bm \theta_m^{(i)}}+ (\mathbf r^{(i)}(\mathbf x^*))^T  (\tilde {\mathbf R}^{(i)})^{-1}(\mathbf y^F- \mathbf f^M(\mathbf x_{1:n}, {\bm \theta}^{(i)})-\mathbf H {\bm \theta_m^{(i)}} ), \]
where $\mathbf r^{(i)}(\mathbf x^*)$  and ${\mathbf {\tilde R}}^{(i)}$ are obtained by plugging in the $i$th  posterior samples. The  predictive interval can also be computed  by specifying the argument \code{interval\_est} in \code{predict.rcalibration} function.

Code below implements the GaSP calibration of parameter estimation and predictions for the example in \citep{bayarri2007framework} that was discussed in the previous subsection. 
\begin{example}
R> m_gasp=rcalibration(input, output,math_model = Bayarri_07,theta_range = theta_range,
+                     X =X, have_trend = T,discrepancy_type = 'GaSP')
R> m_gasp_pred=predict(m_gasp,testing_input,math_model=Bayarri_07,
+                     interval_est=c(0.025, 0.975),X_testing=X_testing) 
\end{example}

Predictions and posterior samples from the GaSP calibration are plotted in Figure  \ref{fig:bayarri_2007}. The associated predictive error is shown in Table \ref{tab:prediction_bayarri_2007}.  Comparing the first two rows in Table \ref{tab:prediction_bayarri_2007},  the predictive RMSE in the no-discrepancy calibration is around $0.25$, and it decreases to $0.15$, after adding the discrepancy function modeled by GaSP. Around 97.5\% of the held-out truth is covered by the 95\%  {predictive interval} of the reality in the GaSP calibration. In comparison, the 95\%  predictive interval of the reality in the no-discrepancy calibration only covers around $80\%$ of the held-out reality.

As shown in the right panel in Figure \ref{fig:bayarri_2007},  the posterior samples from GaSP  spread over a large range, reflecting the large uncertainty in  parameter estimation. This is because the discrepancy modeled by GaSP with the frequently used kernel listed in Table \ref{tab:kernel} is very flexible. As a result, the calibrated computer model by GaSP can be less accurate in predicting the reality. To address this problem, we introduce a new approach that induces a scaled kernel to constrain the discrepancy function.

\begin{table}[t]
\begin{center}
\begin{tabular}{lrrrrr}
  \hline
 & RMSE (CM+trend) &RMSE (with discrepancy)&$P_{CI}(95\%)$ & $L_{CI}(95\%)$    \\
  \hline
  GaSP           &{0.253}&0.151 &{0.975}&{0.880}  \\
  S-GaSP           &{0.228}& 0.131&{0.955} &1.15  \\
  No-discrepancy           &{0.250}&/&{0.795}  &0.409 \\
  \hline
\end{tabular}
\end{center}
   \caption{Predictive accuracy and uncertainty assessment. The RMSE of the out-of-sample prediction based on the calibrated  computer model (CM) and trend are shown in the second column. The predictive RMSE by the summation of the calibrated  computer model, trend and discrepancy is given in the third column. The proportion of the held-out reality covered in the $95\%$ predictive interval, and average lengths of predictive intervals are given in the last two columns, respectively.    }

   \label{tab:prediction_bayarri_2007}
\end{table}

\subsection{Scaled Gaussian stochastic process models of discrepancy functions}
Scaling the kernel of GaSP for model calibration was introduced in \cite{gu2018sgasp}, where the  random $L_2$ distance between the discrepancy model was scaled to have more prior probability mass at small values, as small distance  indicates the computer model fits the reality well. Note that we leave $\sigma^2$ as a free parameter to be estimated from data, and thus S-GaSP is still a flexible model of discrepancy.

In \CRANpkg{RobustCalibration}, we implemented the discretized  S-GaSP  to scale the random mean squared error between the reality and computer model: 
\begin{align}
&\delta_{z_d}(\mathbf x) = \left\{ \delta(\mathbf x) \mid \frac{1}{n}{\sum^{n}_{i=1} \delta( \mathbf x_{i} )^2} =Z_d\right\}, 
\label{equ:delta_z_d}
\end{align}
with the subscript `d' denoting discretization, and the density of $Z_d$ is defined as   
\begin{align}
p_{Z_d}(z) &=	\frac{g_{Z_d}\left(z  \right) p_{ \delta}\left(Z_d=z  \right)}{\int_0^\infty g_{Z_d}\left( t \right)  p_{ \delta}\left(Z_d = t \right)d t },
\label{equ:p_Z_d}
\end{align}
where $p_{ \delta}(Z_d = z)$ is the density of $Z_d$ induced by the GaSP with kernel $K(\cdot,\cdot)$, and $g_Z(\cdot)$ is a nondecreasing function that places more probability mass on smaller values of the $L_2$ loss.   For a frequently used kernel function in Table \ref{tab:kernel}, the probability measure  of the GaSP places a large probability at large $L_2$ loss $p_{ \delta}\left(Z_d=z  \right)$  when the correlation is large, dragging the calibrated computer model away from the reality. In the S-GaSP calibration, the measure for $Z_d$ was scaled to have more probability mass near zero by a scaling function $g_{Z_d}(z)$. The default scaling function is assumed to be an exponential distribution,
\begin{equation}
 g_{Z_d}(z) = \frac{\lambda_z }{2\sigma^2 } \exp\left(-\frac {\lambda_z z}{2\sigma^2 }\right),
 \label{equ:g_Z_d}
 \end{equation}
where $\lambda_z$ controls how large the scaling factor is. Large $\lambda_z$ concentrates more prior probability mass at the origin. As shown in \cite{gu2022theoretical}, $\lambda_z \propto \sqrt{n}$ guarantees the convergence of the calibration parameters to the ones that minimize the $L_2$ distance between the reality and computer model. 
We let  the default choice be $\lambda_z=(\lambda || \bm {\tilde \gamma}||)^{-1/2}$, where $\lambda=(\sigma^2_0/\sigma^2n)$ is the regularization parameter in the kernel ridge regression, and $\tilde {\bm \gamma}=(\gamma_1/L_1,...,\gamma_{p_x}/L_{p_x})^T$,  with $L_i$ being the length of domain of the $i$th coordinate of the calibration parameter. We allow users to specify $\lambda_z$ via the argument  \code{lambda\_z} in the \code{rcalibration} function. 

The default choice of the scaling function in (\ref{equ:g_Z_d})  has computational advantages.  After marginalizing out $Z_d$, it follows from  Lemma 2.4 in \cite{gu2018sgasp} that $\delta_{z_d}(\cdot)$  has the covariance function  
\begin{equation}
\sigma^2 K_{z_d}(\mathbf x_a, \mathbf x_b)=\sigma^2 \left\{K(\mathbf x_a,\mathbf x_b ) - \mathbf r^T(\mathbf x_a)  \left(\mathbf R+\frac{n\mathbf I_n}{\lambda_z} \right)^{-1} \mathbf r(\mathbf x_b)\right\},
\label{equ:sigma_2_K_z_d}
\end{equation}
for any $\mathbf x_a, \mathbf x_b $,  where $\mathbf r(\mathbf x)=(K(\mathbf x, \mathbf x_1),...,K(\mathbf x, \mathbf x_n))^T$ for any $\mathbf x$. The covariance matrix of the S-GaSP model of $(\delta_z(\mathbf x_1),..., \delta_z(\mathbf x_n))^T$ follows
\begin{equation}
\sigma^2 \mathbf R_z=\sigma^2  \left\{\mathbf R- \mathbf R   \left(\mathbf R+\frac{n\mathbf I_n}{\lambda_z} \right)^{-1}\mathbf R\right\}.    
 \label{equ:R_z}
 \end{equation}
 The closed-form expression in Equation (\ref{equ:sigma_2_K_z_d}) avoids sampling $(\delta_z(\mathbf x_1),...,\delta_z(\mathbf x_n))$ to obtain $Z_d$ from the posterior distribution, which greatly improves the computational speed.

For any covariance function $K(\cdot,\cdot)$ in the GaSP calibration, a scaled kernel was induced in S-GaSP calibration, shown in the last row of Table \ref{tab:kernel}. The S-GaSP model of the discrepancy can be specified by the argument \code{discrepancy\_type='S-GaSP'} in the \code{rcalibration} function. Similar to the model of GaSP-calibration, we also provide MLE and posterior samples for estimating the parameters, with arguments  \code{method='mle'} and \code{method='post\_sample'}, respectively.  The MLE of parameters in S-GaSP are computed by maximizing the profile likelihood 
\begin{equation}
(\hat {\bm \theta}, \hat {\bm \gamma}, \hat \eta ) =\argmax_{(\bm \theta, \bm \gamma, \eta)} \mathcal L_{K_{Z_d}}( \bm \theta, \bm \gamma, \eta, \mid { \hat {\bm \theta}_m},\hat \sigma^2_0 ),   
\label{equ:MLE_SGaSP} 
\end{equation}
where  $({ \hat {\bm \theta}_m}, \hat \sigma^2_0)$ denotes the MLE  of trend  and noise variance parameters in S-GaSP calibration, and $\mathcal L_{K_{Z_d}}$ is the profile likelihood with respect to the scaled kernel $K_{Z_d}$ in (\ref{equ:sigma_2_K_z_d}). Furthermore, point predictions and intervals can be obtained by the \code{predict.rcalibration} function as well. 

{The code below gives S-GaSP calibration and prediction of the example in \cite{bayarri2007framework}}. 
\begin{example}
R> m_sgasp=rcalibration(input, output,math_model = Bayarri_07,theta_range = theta_range,
+                      X =X, have_trend = T,discrepancy_type = 'S-GaSP')
R> m_sgasp_pred=predict(m_sgasp,testing_input,math_model=Bayarri_07,
+                      interval_est=c(0.025, 0.975),X_testing=X_testing) 
\end{example}

Predictions from S-GaSP are plotted as blue curves in Figure  \ref{fig:bayarri_2007} and numerical comparisons are given in Table \ref{tab:prediction_bayarri_2007}. With the discrepancy modeled by S-GaSP, the prediction of reality is more accurate at $x \in [0.5,1.5]$, compared to no-discrepancy calibration.  Note that predictions from the S-GaSP model also extrapolate the reality at $x\in [3,5]$ accurately, because the calibrated computer model is  closer to the truth.  Furthermore, the 95\% predictive interval by S-GaSP covers around $95\%$ of the held-out samples.

{To compare and illustrate different calibration approaches for  differential equations}, we discuss another example of  the Lorenz-96 system used in modeling atmospheric dynamics \citep{lorenz1996predictability}. The mathematical model is written as the following differential equations:  
\begin{align}
\dot x_j(t) = (x_{j+1}(t)-x_{j-2}(t))x_{j-1}(t) - x_j(t)  + \theta, 
 \end{align}
 for $j=1,...,k$ states, with $k=40$ and $\theta$ is a scalar value of the force. We further denote that $x_{-1}=x_{k-1}$, $x_0 = x_k$ and $x_{k+1}=x_{1}$ for any $t$. The latent variable $x_j$ can model the atmospheric quantities, such as temperature or pressure, measured at $k$ positions along a constant latitude circle. This model is widely used for data assimilation \citep{maclean2020surrogate,brajard2020combining}.
 
We consider  model calibration in two scenarios:  
 \begin{align}
 \mbox{Scenario 1: } \quad \quad \quad \quad \quad \quad \quad \quad \quad \quad  y_j(t)&= x_j(t) +\epsilon,    \quad \quad \quad \quad \quad \quad \quad   \quad \quad \quad \quad \quad  \label{equ:no_discrepancy}\\ 
\mbox{Scenario 2: } \quad \quad \quad \quad \quad \quad \quad \quad \quad \quad y_j(t)&= x_j(t) +2t \sin\left(\frac{2\pi j}{k}\right)+ \epsilon,   \quad \quad \quad \quad \quad \quad \quad \quad \quad \quad \quad \quad \label{equ:with_discrepancy}
 \end{align}
 for $j=1,...,k$ and  $\epsilon \sim \mathcal{N}(0, 1)$ being an independent Gaussian noise. In the first scenario, the mathematical model is the true model and in the second scenario, a discrepancy term is included for simulating the field data. The discrepancy is treated as unknown. We initialize the states by a multivariate normal {distribution} with the covariance generated from a Wishart distribution with $k$ degrees of freedom and {the scale matrix being a diagonal matrix}. We use the Runge Kutta method with order 4 to numerically solve the system. Since the computer model is fast,  we do not include an emulator. We simulate the reality at $40$ time points, with a time step of $0.05$ between the time points. 
    \begin{figure}[t]
\centering
  \begin{tabular}{ccc}

	\hspace{-.1in} \includegraphics[height=0.5\textwidth,width=0.33\textwidth]{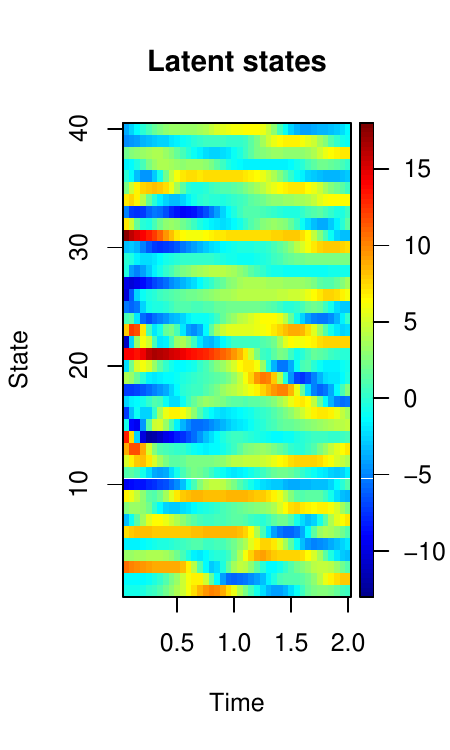}
	\hspace{-.1in} \includegraphics[height=0.5\textwidth,width=0.33\textwidth]{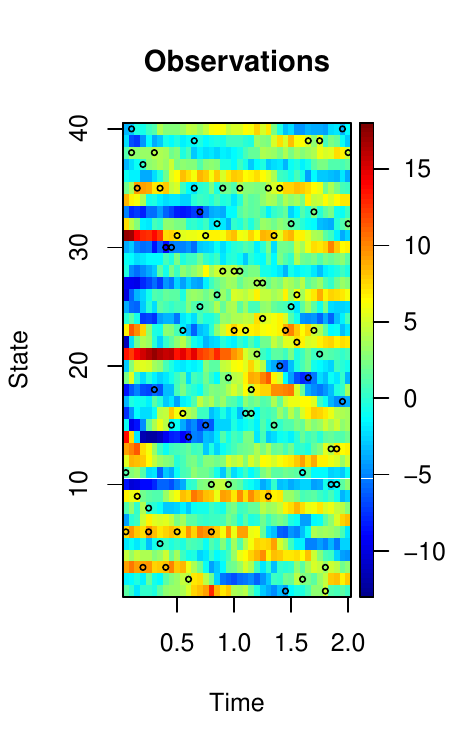}
		\hspace{-.1in} \includegraphics[height=0.5\textwidth,width=0.33\textwidth]{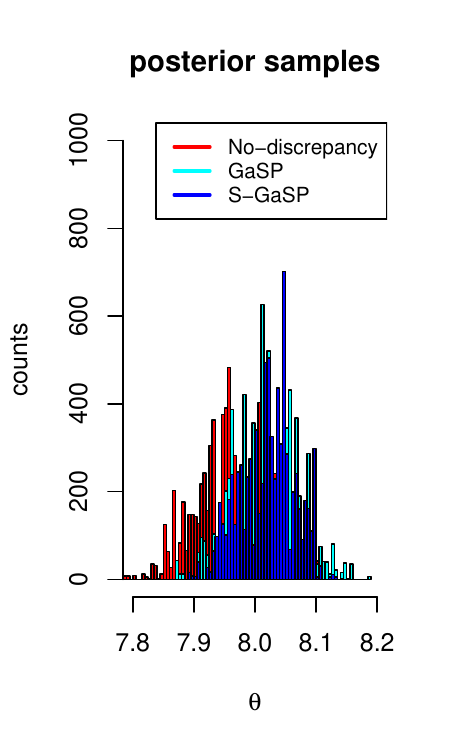} \vspace{-.3in} \\
	\hspace{-.1in} \includegraphics[height=0.5\textwidth,width=0.33\textwidth]{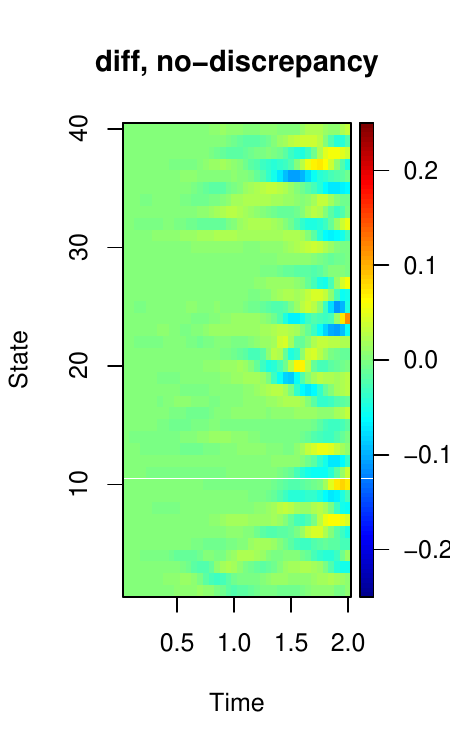}
	\hspace{-.1in} \includegraphics[height=0.5\textwidth,width=0.33\textwidth]{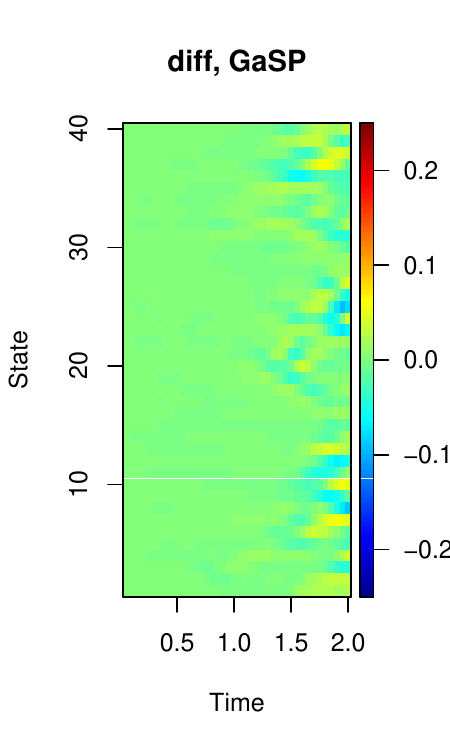}
		\hspace{-.1in} \includegraphics[height=0.5\textwidth,width=0.33\textwidth]{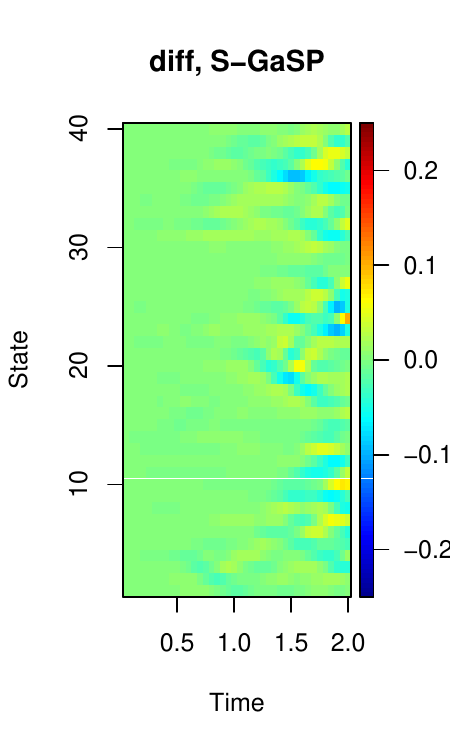} \vspace{-.15in}

\end{tabular}
   \caption{The reality and full observations of Lorenz-96 system in Scenario 1 are plotted in the upper left panel and upper middle panel, respectively, where the black circles are the $5\%$ observations used for model calibration.  Posterior samples of calibration parameters of different calibration methods are plotted in the upper right panel.  The lower panels give the  difference between the reality and calibrated computer model  of all latent states.   }
 \label{fig:lorenz_96}
\end{figure}

  \begin{figure}[t]
\centering
  \begin{tabular}{ccc}

	\hspace{-.1in} \includegraphics[height=0.5\textwidth,width=0.33\textwidth]{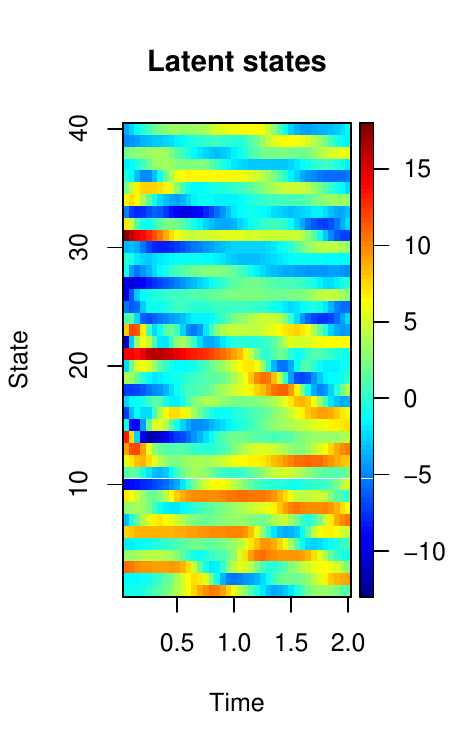}
	\hspace{-.1in} \includegraphics[height=0.5\textwidth,width=0.33\textwidth]{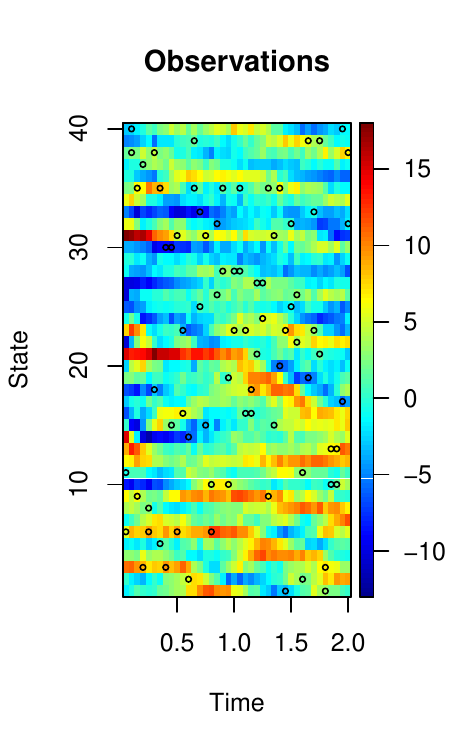}
		\hspace{-.1in} \includegraphics[height=0.5\textwidth,width=0.33\textwidth]{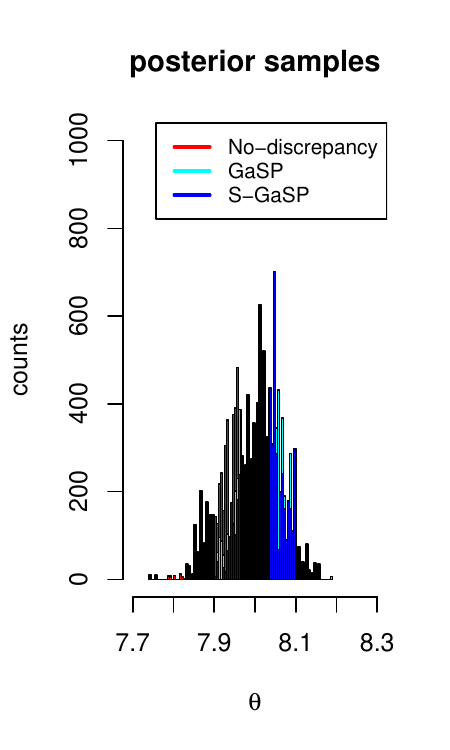} \vspace{-.3in} \\
	\hspace{-.1in} \includegraphics[height=0.5\textwidth,width=0.33\textwidth]{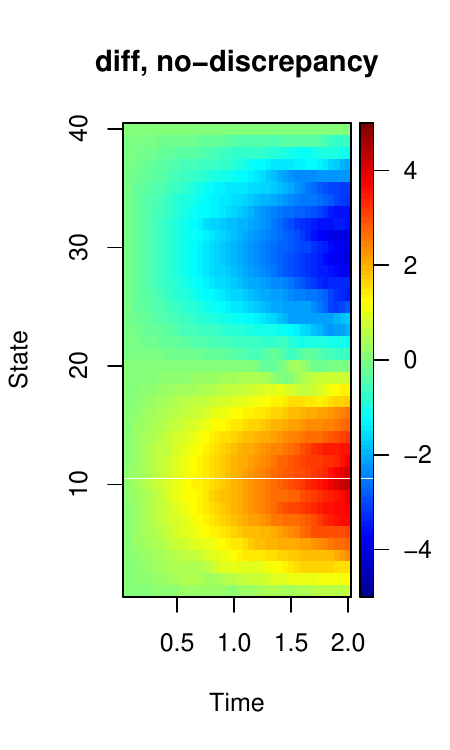}
	\hspace{-.1in} \includegraphics[height=0.5\textwidth,width=0.33\textwidth]{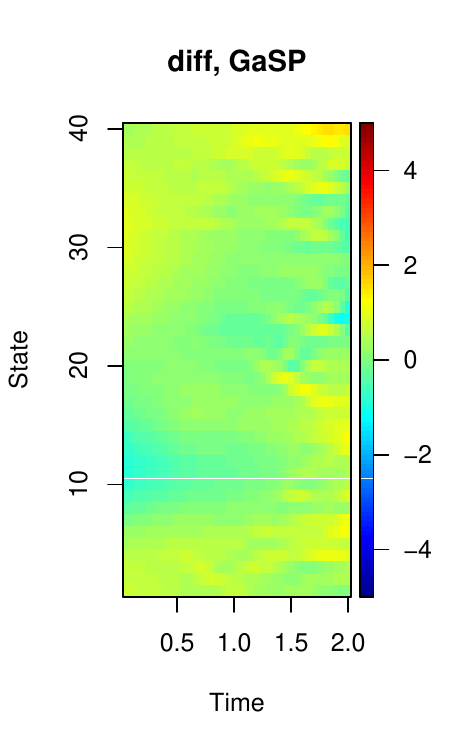}
		\hspace{-.1in} \includegraphics[height=0.5\textwidth,width=0.33\textwidth]{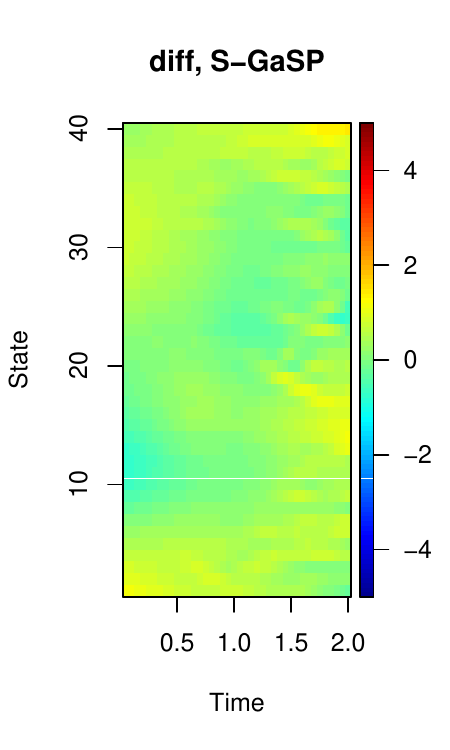} \vspace{-.15in}

\end{tabular}
   \caption{The reality of the discrepancy-included Lorenz-96 system (Scenario 2) is plotted in the upper left panel. The simulated observations are plotted in the upper middle panel, where the $5\%$ of the observations plotted as black circles are used in model calibration.  Posterior samples of calibration parameters of different calibration methods are plotted in the upper right panel.  The lower panels show the differences between the reality and the predictive mean  of all latent states.   }
 \label{fig:lorenz_96_discrepancy}
\end{figure}

 The unobserved reality simulated by the Lorenz-96 system is plotted in upper  panel in Figure \ref{fig:lorenz_96}.  Only $5\%$ observations  (i.e. 2 observations at each time point, plotted as black circles in upper middle panel) are available for model calibration. The posterior parameters of no-discrepancy calibration, GaSP and S-GaSP calibration are plotted in the upper right panel. Since the computer model (Lorenz-96) is the true model {in this scenario,} the no-discrepancy calibration is expected to perform well. Even though one allows a flexible discrepancy function, modeled as GaSP or S-GaSP, it seems parameters are estimated reasonably well in both approaches.  In all methods, posterior samples are close to the true parameter ($\theta=8$), and the range of the posterior samples is  narrow compared to the parameter range $[-20, 20]^2$. Furthermore, the estimate states by the calibrated computer model and the reality are plotted in the lower panels in Figure \ref{fig:lorenz_96}. All methods have small estimation errors, as the parameters are estimated well.

The reality, observations, posterior samples, and predictions from the different calibration models of a discrepancy-included Lorenz-96 system are graphed in Figure \ref{fig:lorenz_96_discrepancy}. Note that since the discrepancy is included, there is no true calibration parameter.  The states estimated by the no-discrepancy calibration model (shown in the lower left panel) have {a} relatively large error. This is not surprising as the Lorenz-96 system is a misspecified model. The predictive means of GaSP and S-GaSP models which includes both the calibrated computer model and discrepancy function is more accurate, as both models capture the discrepancy between the reality and computer model.  The estimation error of all models is larger than the ones when there is no discrepancy. This is because the measurement error is large and we only observe 2 states at each time point. Increasing the number of observations or reducing the variance of measurement error can improve predictive accuracy. 

\subsection{Calibration with repeated experiments}
Repeated experiments or replicates are commonly used in experiments, {as they are } helpful for assessing the experimental error. Consider, for example, $k_i$ replicates  available for each observable input $\mathbf x_i$, denoted as $\mathbf y^F_i=(y_1^F(\mathbf x_i),...,y^F_{k_i}(\mathbf x_i))^T$. Denote $\bar{\mathbf y}^F=(\sum^{k_1}_{j=1}y^F_{j}(\mathbf x_1)/k_1,...,\sum^{k_n}_{j=1}y^F_{j}(\mathbf x_n)/k_n)^T$ the aggregated data. The probability density of the field observations can be written as: 
\begin{align*}
p(\mathbf y^F_1,...,  \mathbf y^F_n \mid \bm \mu, \bm \delta, \bm \theta,  \sigma^2_0) &= (2\pi \sigma^2_0)^{-\frac{\sum^n_{i=1} k_i}{2}} \exp\left(-\frac{\sum^n_{i=1} \omega_i \sum^{k_i}_{j=1}  (y^F_j(\mathbf x_i)- f^M(\mathbf x_i,\bm \theta)-\mu_i-\delta(\mathbf x_i) )^2  }{2\sigma^2_0}\right) \\
&=   (2\pi\sigma^2_0)^{-\frac{\sum^n_{i=1} k_i}{2}} \exp\left(-\frac{ S^2_f }{2\sigma^2_0}- \frac{\sum^n_{i=1} k_i \omega_i ( \bar{ y}^F_i-f^M(\mathbf x_i,\bm \theta)-\mu_i-\delta(\mathbf x_i) )^2 }{2\sigma^2_0}\right)
\end{align*}
where $S^2_f=\sum^n_{i=1}\omega_i\sum^{k_i}_{j=1}   (y^F_j(\mathbf x_i)-\bar{ y}^F_i )^2=\sum^n_{i=1}{\omega_i(\mathbf y^F_i-\bar{ y}^F_i\mathbf 1_{k_i} )^T(\mathbf y^F_i-\bar{ y}^F_i \mathbf 1_{k_i}) }$ with the subscript `f' denoting the field observations. After integrating out the discrepancy term, assumed to be modeled as a GaSP as an example, the marginal likelihood of the parameters follows
\begin{align}
\mathcal L_K(\bm \theta, \bm \theta_m, \sigma^2_0, \bm \gamma,\eta )\propto (\sigma^2_0)^{-\frac{\sum^n_{i=1} k_i}{2} }| \tilde {\tilde {\bm R}}|^{-\frac{1}{2} }  \exp\left\{- \frac{(\bar{\mathbf y}^F-\mathbf f^M_{\theta}-\mu \mathbf 1_n )^T \tilde {\tilde {\bm R}}^{-1} (\bar{\mathbf y}^F- \mathbf f^M_{\theta}-\mu \mathbf 1_n)}{2 \sigma^2_0} -\frac{S^2_f}{2\sigma^2_0} \right\}, 
\label{equ:lik_replicates}
\end{align} 
where $\tilde {\tilde {\bm R}}= ( \mathbf R/\eta+  \tilde {\bm \Lambda})$, with $\tilde {\bm \Lambda}$ being an diagonal matrix with diagonal entry $\Lambda_{ii}=1/(\omega_ik_i)$, for $i=1,2,...,n$, and $\eta= \sigma^2_0/\sigma^2$. The S-GaSP calibration with replications can also be defined, by simply replacing $\mathbf R$ by $\mathbf R_z$ where the $(i,j)$th term is defined by the scaled kernel in Table \ref{tab:kernel}. 

The likelihood in (\ref{equ:lik_replicates})  has a clear computational advantage. If one directly computes the covariance and its inverse, the computational complexity is $O( (\sum^{n}_{i=1}k_i)^3 )$, whereas the computational order by Equation (\ref{equ:lik_replicates}) is only $O(n^3)+O(\sum^{n}_{i=1}k_i)$. Suppose $k_i=k$, for $i=1,...,n$, using Equation (\ref{equ:lik_replicates}) is around $k^3$ times faster than directly computing the likelihood, with no loss of information.

 One can specify replications by inputting a $n\times k$ matrix into the argument \code{observations} in the \code{rcalibration} function for scenarios with $n$ observable inputs, each containing $k$ repeated measurements. Or one can give a list of \code{observations} in the \code{rcalibration} function with size $n$, where each element in the list contains $k_i$ values of replicates, for $i=1,...,n$. For observations with replicates,  MLE and posterior sampling were implemented in the \code{rcalibration} function, and predictions were implemented in the \code{predict.rcalibration} function.

\subsection{Statistical emulators}
Users can specify a mathematical model via the  \code{math\_model} argument in \code{rcalibration} function if the closed-form expression of the mathematical model is available. However, closed-form solutions of physical systems are typically unavailable, and numerical solvers are required. Simulating the outputs from computer models can be slow. For instance, the TITAN2D computer model, which simulates pyroclastic flows for geological hazard quantification, takes roughly 8 minutes per set of inputs  \citep{simakov2019modernizing}. As hundreds of thousands of model runs are sometimes required for model calibration \citep{anderson2019magma}, directly computing the computer model by numerical methods is often prohibitive. 
In this scenario, we construct a statistical emulator as a surrogate model, to approximate the computer simulation, based on a small number of computer model runs. 

To start, assume that we have obtained the output of the computer model at $D$ input design points, $(\bm \theta_1,...,\bm \theta_D)$, often selected from an ``space-filling" design, such as the Latin hypercube design \citep{santner2003design}. There are two types of computer model outputs: 1) scalar-valued outputs: $f^M(\bm \theta) \in \mathbb R$, and 2) vector-valued outputs at $k$ coordinates $\mathbf f^M(\mathbf x_{1:k}, \bm \theta) \in \mathbb R^k$, where $k$ can be as large as  $10^6$. 
Various packages are available for fitting scalar-valued GP emulators, such as \CRANpkg{DiceKriging} (\cite{roustant2012dicekriging}), \CRANpkg{GPfit} (\cite{macdonald2015gpfit}), and \CRANpkg{RobustGaSP} (\cite{gu2018robustgasp}). Some of these packages were used in Bayesian model calibration packages \citep{palomo2015save,carmassi2018calico}. However, the emulator of computer models with vector-valued outputs, a common scenario in different disciplines \citep{higdon2008computer,ma2022multifidelity,li2022efficient,fang2022reliable} was rarely implemented in the model calibration packages.

 In \CRANpkg{Robustcalibration}, we call the \code{rgasp} function and \code{ppgasp} function from the \CRANpkg{RobustGaSP} package for emulating scalar-valued and vector-valued computer model outputs, respectively. The parallel partial Gaussian process (PP-GaSP) by the \code{ppgasp} function has two advantages. First, computing the predictive mean by  the PP-GaSP only takes $O(kD)+O(D^3)$  operations, which is particularly suitable for computer models with a large number of outputs ($k$). 
 {Second, as the covariance over $\bm \theta$ is shared across all grids, the estimation of PP-GaSP is more stable than building a separate emulator on each output coordinate.} Furthermore,  the marginal posterior mode with robust parameterization typically avoids the degenerated estimation of the covariance matrix (\cite{gu2019jointly}). 

To call the emulator, users can select the argument \code{simulator=0}, and then specify the simulation runs in arguments \code{input\_simul}, and  \code{output\_simul}. One can input a vector into \code{output\_simul} to emulate scalar-valued computer models, or a matrix into \code{output\_simul} to call the PP-GaSP emulator for vector-valued outputs. For both scenarios, 
 we use a modular approach to fit the PP-GaSP emulator  \citep{bayarri2007framework,liu2009modularization}, where  the predictive distribution of the emulator depends on simulator runs, but not on field observations. After fitting an emulator, the predictive mean of the PP-GaSP emulator will be used to approximate the computer model at any unobserved $\bm \theta^*$.

We illustrate the efficiency and accuracy of the PP-GaSP emulator for approximating  differential equations from \cite{box1956application}, where the interaction between  two chemical substances $y_1$ and $y_2$ are modeled as 
\begin{align*}
\dot{y}_1(t)&=10^{\theta_1-3} y_1(t), \\
\dot{y}_2(t)&=10^{\theta_1-3} y_1(t) - 10^{\theta_2-3}{y}_2(t).
\end{align*}
In each of the 6  time points $t=10,20,40,80,160$, and $320$, 2 replicates of the second chemical substance are available in \cite{box1956application}. Initial conditions are $y_{1}(t=0)=100$ and $y_2(t=0)=0$. The computer model contains two parameters $\theta_1\in [0.5, 1.5]$ and $\theta_2\in [0.5,1.5]$. 

To start, we first use the default method in the \code{ode} function from the \CRANpkg{deSolve} package \citep{soetaert2010solving} to numerically solve the ODEs at a given initial condition and parameter set: 
\begin{example}
library(deSolve)
R> Box_model <- function(time, state, parameters) {
+   par <- as.list(c(state, parameters))
+   with(par, {
+     dM1=-10^{parameters[1]-3}*M1
+     dM2=10^{parameters[1]-3}*M1-10^{parameters[2]-3}*M2
+     list(c(dM1, dM2))
+   })
+ }
R> Box_model_solved<-function(input, theta){
+   init <- c(M1 = 100 , M2 = 0)
+   out <- ode(y = init, times = c(0,input), func = Box_model, parms = theta)
+   return(out[-1,3])
+ }
\end{example}

We  specify the observations and  range of calibration parameters from \cite{box1956application}  below. 
\begin{example}
R> n=6
R> output=t(matrix(c(19.2,14,14.4,24,42.3,30.8,42.1,40.5,40.7,46.4,27.1,22.3),2,n))
R> input=c(10,20,40,80,160,320)
R> theta_range=matrix(c(0.5,0.5,1.5,1.5),2,2)
###the testing input for emulator should contain the observed inputs
R> testing_input=as.matrix(seq(1,350,1)) 
#if observed inputs are not included, then add it
R> set_diff_obs=setdiff(input,testing_input) 
R> testing_input=sort(c(testing_input,set_diff_obs)) 
\end{example}

We compare model calibration and predictions based on the numerical solver  by the \CRANpkg{deSolve} package, and by the GaSP emulator approach below. We first implement the direct approach, where the numerical solver is called {to generate posterior samples}. 
\begin{example}
R> m_sgasp_time=system.time({
+ m_sgasp=rcalibration(input,output,math_model=Box_model_solved,
+                     sd_proposal=c(0.25,0.25,1,1),
+                     theta_range=theta_range)
+ m_sgasp_pred=predict(m_sgasp,testing_input,math_model=Box_model_solved,
+                           interval_est=c(0.025,0.975))})
\end{example}

We then generate 50 design points from maximin Latin hypercube design by the \CRANpkg{lhs} package, and run a numerical solver at these design points. {As the simulator outputs a vector $\mathbf f^M(t_{1:k},\bm \theta) \in \mathbb R^k$ for calibration parameter, we call the PP-GaSP emulator} to predict the output at all time points.  The simulator runs are then specified as \code{input\_simul} and \code{output\_simul} in the \code{rcalibration} function. We let \code{simul\_type=0} to call the emulator instead of the numerical solver {to generate posterior samples}. {The \code{loc\_index\_emulator} below gives  a subset of the output coordinates for the   the PP-GaSP emulator to be predicted and by default, the function will predict all output coordinates.}  
\begin{example}
R> m_sgasp_emulator_time=system.time({
+   ##constructing simulation data
+   D=50
+   p=2
+   lhs_sample=maximinLHS(n=D,k=p)
+   input_simul=matrix(NA,D,p)
+   input_simul[,1]=lhs_sample[,1]+0.5 
+   input_simul[,2]=lhs_sample[,2]+0.5 
+   k_simul=length(testing_input)
+   
+   output_simul=matrix(NA,D,k_simul)
+   for(i_D in 1:D){
+    output_simul[i_D,]=Box_model_solved(c(testing_input), input_simul[i_D,])
+   }
+   ##create loc_index_emulator as a subset of the simulation output
+   loc_index_emulator=rep(NA,n)
+   for(i in 1:n){
+     loc_index_emulator[i]=which(testing_input==input[i])
+   }
+   
+   ##emulator
+   m_sgasp_with_emulator=rcalibration(input,output,simul_type=0,
+                       input_simul=input_simul, output_simul=output_simul,simul_nug=T,
+                       loc_index_emulator=loc_index_emulator,
+                       sd_proposal=c(0.25,0.25,1,1),
+                       theta_range=theta_range)
+   m_sgasp_with_emulator_pred=predict(m_sgasp_with_emulator,testing_input)
+   })
R> m_sgasp_time[3]/m_sgasp_emulator_time[3]
elapsed 
6.252316 
\end{example}
 The approach with an emulator only costs around $1/6$ of the computational time in this example, even though the numerical solver of this simple ODE is fast. For computer models that take a few  or hours to run, the emulator approach can {substantially reduce} the computational cost. 
 
  \begin{figure}[t]
\centering
  \begin{tabular}{ccc}

	\hspace{-.2in} \includegraphics[scale=.63]{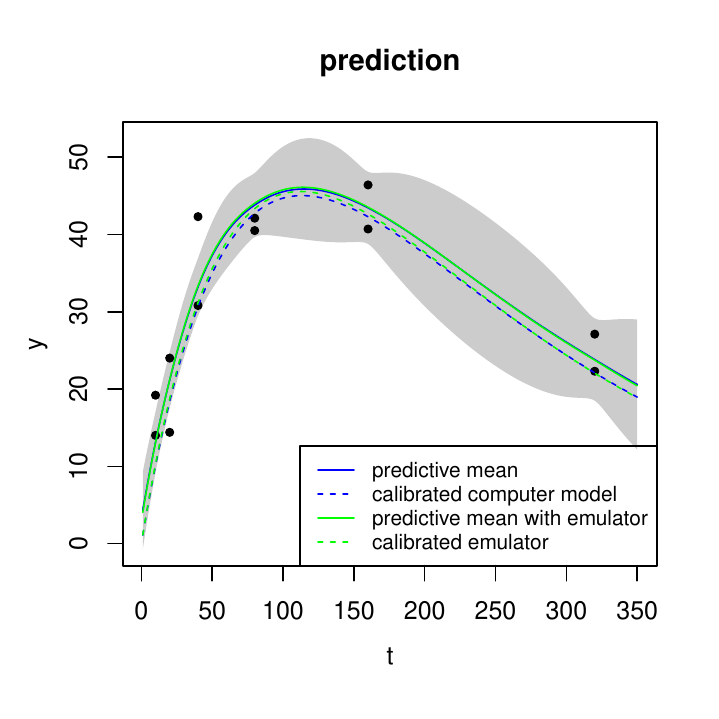}
	\hspace{-.3in} \includegraphics[scale=.63]{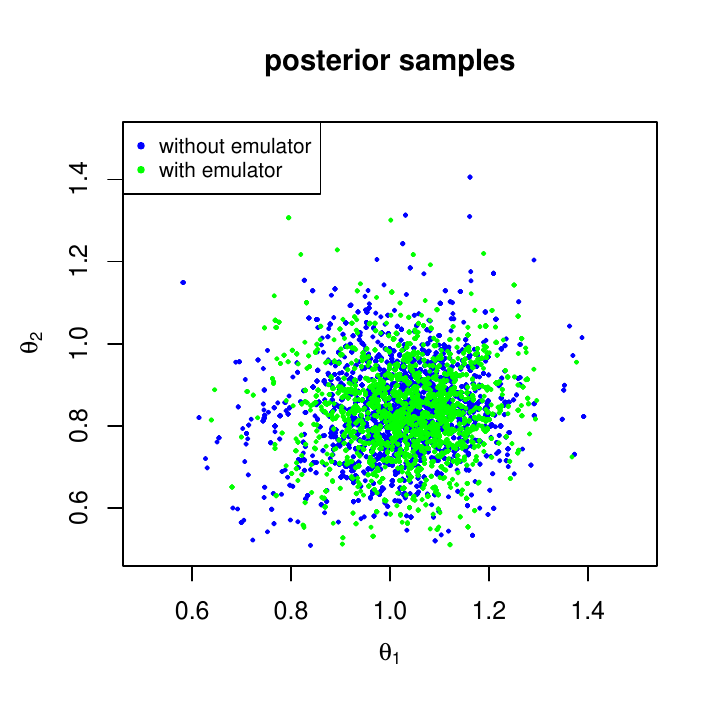}
\vspace{-.2in}
\end{tabular}
   \caption{Comparison between the Bayesian model calibration based on the numerical solver and the PP-GaSP emulator of the computer model. In the left panel, the green solid curve is the predictive mean from the calibrated computer model and the discrepancy function, and the {blue} dashed curves are predictions from the calibrated computer model alone, both of which call the numerical solver for each posterior sample. The same result based on the emulator is plotted as  the {green}  curves. The black dots are field observations and the shaded area is the  $95\%$ predictive interval of the reality by the approach that uses the PP-GaSP emulator. In the right panel,   posterior samples based on the numerical solver and the PP-GaSP emulator are denoted by the {blue}  dots and {green}   dots, respectively.    }
 \label{fig:box_model}
\end{figure}

Figure \ref{fig:box_model} gives predictions and posterior samples based on a numerical solver and the PP-GaSP emulator. First, predictions from the calibrated computer model and the discrepancy function, and by the calibrated computer model itself are close to each other, implying that the discrepancy modeled by a S-GaSP is close to be zero. 
Second, predictions and posterior distributions from the numerical solver and  the emulator are close to each other, meaning that the emulator approximates the solver well.  {Calibration with an emulator is preferred as it is faster than numerically solving ODEs each time when generating the posterior samples}. 

\subsection{Calibration with multiple sources of observations}
In reality, one may have observations of different types and they may come from multiple sources.  In \cite{anderson2019magma}, for instance, multiple satellite interferogram and GPS observations measuring the ground deformation during the K\={\i}lauea volcano eruption in Summer 2018 are used for calibrating mechanical models that relate the observations to calibration parameters, such as depth and shape of the magma chamber, magma density, pressure change rate, and so on. In these applications, the measurement bias, such as the atmospheric error, can substantially affect the satellite measurements of the ground deformation \citep{Zebker1997,agram2015noise}. 

 Model calibration using multiple sources of observations was {studied} in \cite{gu2022calibration}. For each source $l$, $l=1,2,...,k$, let the field observations of the $l$ th source at observable input $\mathbf x$ be modeled as
\begin{equation}
\mathbf y^F_l(\mathbf x)=  f^M_l(\mathbf x, \bm \theta)+ \delta(\mathbf x)+ \delta_l(\mathbf x)  +\mu_l+\epsilon_{l},
\label{equ:measurement_bias}
\end{equation}
where  $f^M_l(\mathbf x, \bm \theta)$ is the computer model for source $l$. The discrepancy function is denoted by $\delta(\mathbf x)$ and the source-specific measurement bias is denoted by $\delta_l(\mathbf x)$.
 Measurement bias often appears in the satellite radar interferogram, as atmospheric error could affect the quality of the image.  The  mean  of each data source is denoted by $\mu_l$, and the independent measurement noise is denoted by $\epsilon_{l}$. 

In \CRANpkg{RobustCalibration}, we allow users to integrate observations from multiple sources to calibrate computer models by the function \code{rcalibration\_MS} and to make predictions using the function \code{predict\_MS.rcalibration\_MS}. The posterior sampling method is implemented for parameter estimation for multiple sources of data. Besides, both GaSP  and S-GaSP models can be chosen to model the measurement bias $\delta_l$, for $l=1,...,k$. Similar to calibration model with a single source of data, we allow users to choose a model with no-discrepancy, and with GaSP or S-GaSP model of the discrepancy function $\delta$. Users can choose to have measurement bias or not.  
We also allow users to integrate different types of measurements for model calibration by using different designs of observable inputs.

   To illustrate, we study a synthetic example for calibrating computer models using multiple sources of observations. We assume  the $l$th source of data, $l=1,...,k$, is simulated  below
  \begin{equation}
  y^F_l(x)=sin(\pi x)+ \delta(x)+\delta_l(x)+\epsilon_l(x),
  \label{equ:multi_calibration_measurement_bias}
  \end{equation} 
  where $\delta(.) $ and $\delta(.)$ are independently simulated from Gaussian processes with covariance $\sigma^2K(.,.)$ and $\sigma^2_lK_l(\cdot,\cdot)$, and the independent noise follows $\epsilon_l(x) \overset{i.i.d}{\sim} N(0,\sigma^2_0)$.  We let $\sigma_0=0.05$, $\sigma=0.2$ and $\sigma^2_l=0.5\times l/(l-1)$, for $l=1,...,k$. The  $K(.,.)$ and $K_l(\cdot,\cdot)$ are assumed to follow Mat{\'e}rn kernel with roughness parameter being $\alpha=2.5$, and the range parameter being $\gamma=1/30$ and $\gamma=1/10$, respectively.  We collect $n=100$ observations for each source $l$ with $x$ equally spaced from $[0,1]$. We assume data from $k=5$ sources are available. 
Here the mathematical model follows $f^M_l(x,\theta)=sin(\theta x)$, for $l=1,...,k$, and the reality  is $y^R(x)=f^M_l(x,\theta)+\delta(x)$. The goal is to estimate the calibration parameter ($\theta$), the reality ($y^R(x)=f^M_l(x,\theta)+\delta(x)$), model discrepancy ($\delta(x)$), and measurement bias ($\delta_l(x)$).

We compare four methods. The first two methods are GaSP and S-GaSP  calibration methods to model individual sources of data, with the discrepancy model specified as a GaSP and S-GaSP respectively. The measurement bias terms are modeled by GaSPs.  {For instance, the code below implements S-GaSP model of discrepancy and GaSP model of the measurement bias:}
\begin{example}
> model_sgasp=rcalibration_MS(design=input_measurement,observations=output,p_theta=1,
              math_model=math_model, simul_type=rep(1,length(input_measurement)),
              S=5000,S_0=2000,thinning=1,measurement_bias=T,shared_design=input_model,
              have_measurement_bias_recorded=T,
              discrepancy_type=c(rep('GaSP',length(input_measurement)),'S-GaSP'),
              theta_range=matrix(c(-2*pi,2*pi),1,2),sd_proposal_theta=rep(0.02,2)); 
\end{example}
{where \code{math\_model}, \code{input\_measurement} and \code{output} are lists where each contains the model, input and output of the field observations for each source. The \code{input\_model} is a matrix for the input of the discrepancy function}. 

 {As modeling each source of the data can be time-consuming for some applications, a common solution is to use the aggregated or stacked data $ \bar y^R(x)=\sum^k_{l=1} y^R_l(x)/k$ in calibration, as modeling the aggregated data is around $k$ times faster than modeling the individual sources of data}. Thus, we also include  two methods of modeling the aggregated data for comparison, namely the GaSP Stack and S-GaSP Stack, representing GaSP or S-GaSP discrepancy model, respectively.

    \begin{figure}[t]
\centering
  \begin{tabular}{cc}

	\hspace{-.1in} \includegraphics[height=0.4\textwidth,width=0.5\textwidth]{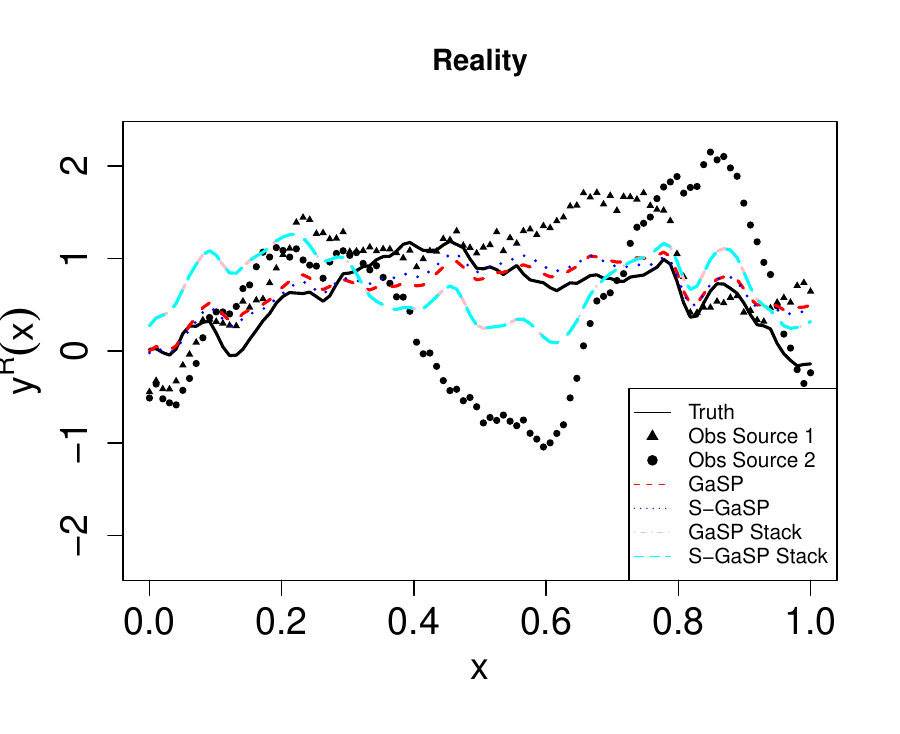}
		\hspace{-.1in} \includegraphics[height=0.4\textwidth,width=0.5\textwidth]{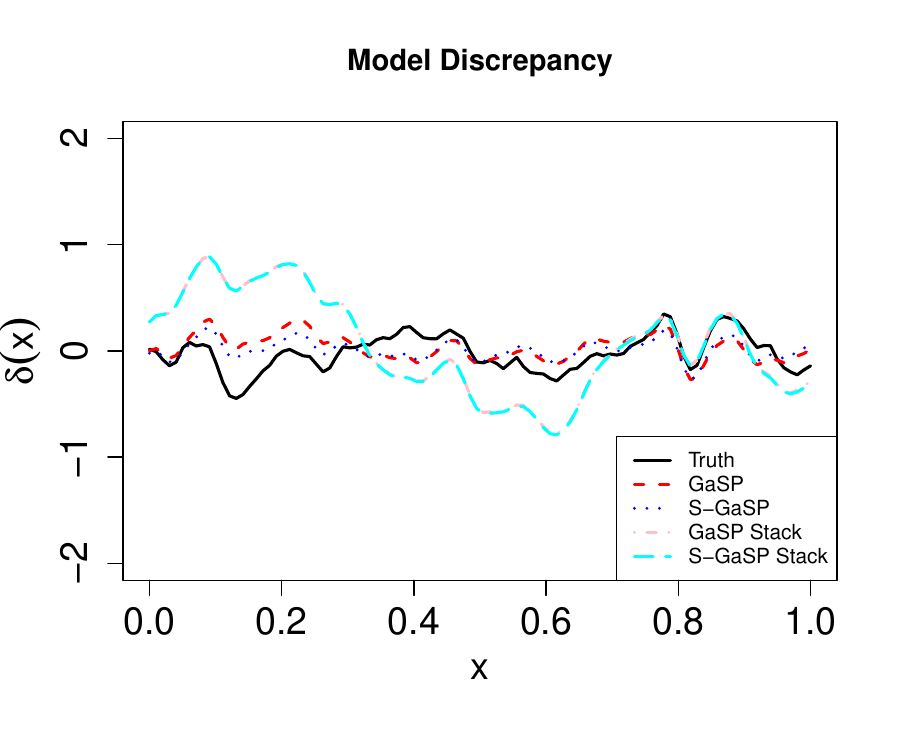} \vspace{-.3in} \\
	\hspace{-.1in} \includegraphics[height=0.4\textwidth,width=0.5\textwidth]{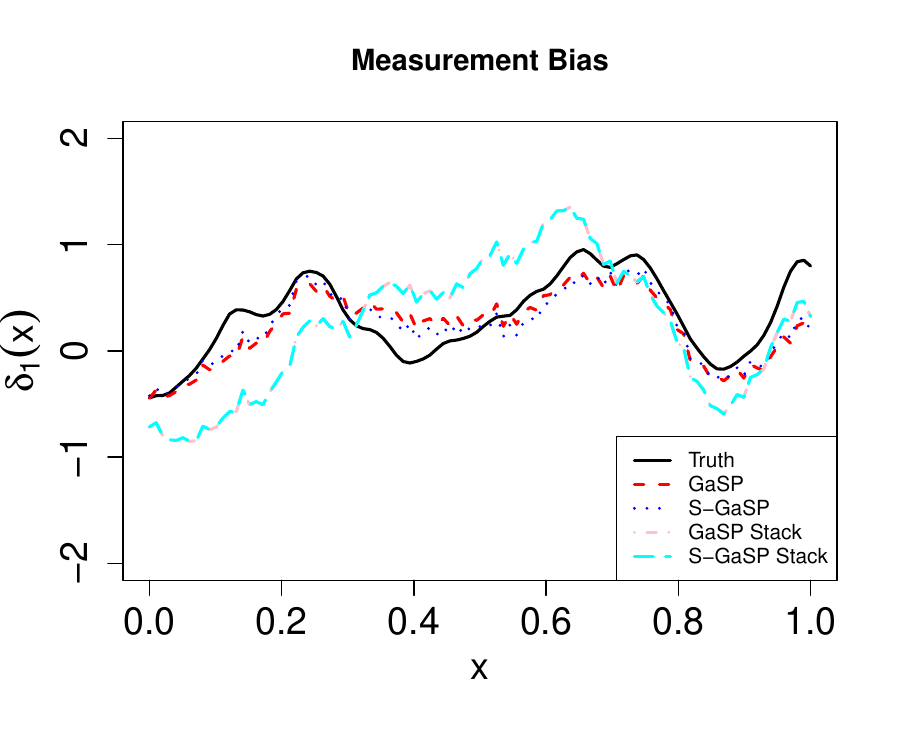}
	\hspace{-.1in} \includegraphics[height=0.4\textwidth,width=0.5\textwidth]{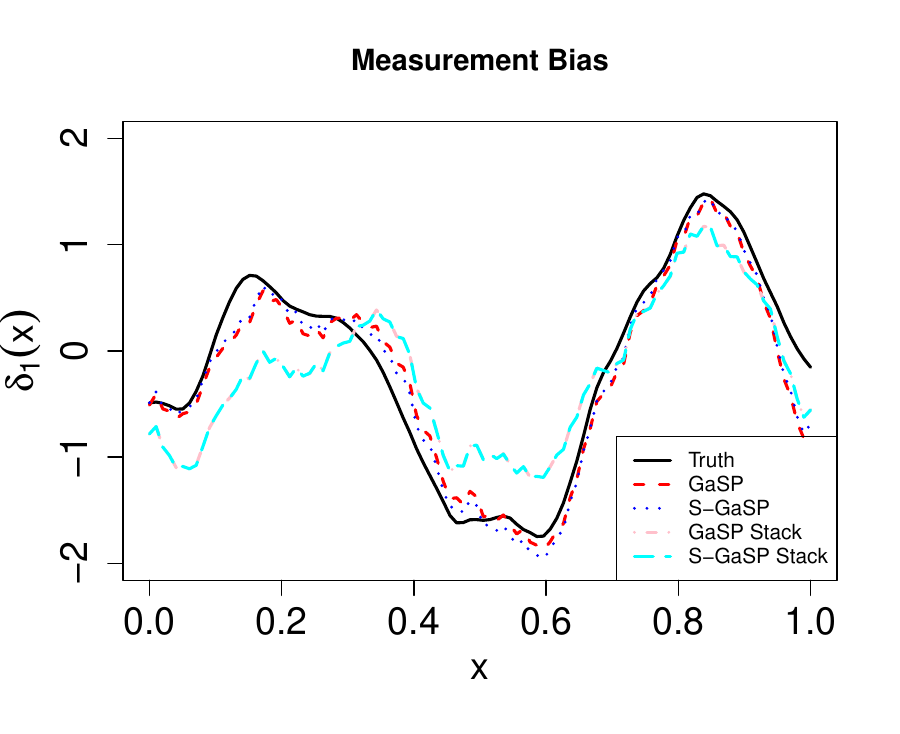}

\end{tabular}
\vspace{-.2in}
   \caption{Comparison between modeling individual data and aggregated data for simulation from  Equation (\ref{equ:multi_calibration_measurement_bias}).  The upper left and right panels give the reality and model discrepancy, whereas the lower two panels give the measurement bias for the first two sources. The estimation results from GaSP, S-GaSP,  GaSP Stack and S-GaSP calibration are plotted as  red, blue, pink and cyan curves. The observations from the first two sources are graphed as the triangles and dots in the upper left panel.  }
 \label{fig:multicalibration_eg}
\end{figure}

In Figure \ref{fig:multicalibration_eg}, we plot the truth, observations of the first two sources of observations, and the estimates of reality, discrepancy, and  measurement bias in two sources from four different methods. Here the GaSP and S-GaSP perform better than the GaSP Stack and S-GaSP Stack calibration methods, as aggregating data leads to loss of information when the data contain measurement bias. The RMSE of reality, discrepancy and measurement bias estimation  of different  methods are given in the Table \ref{tab:prediction_multicalibration}, {which shows the S-GaSP calibration of individual sources of data achieves the highest calibration and predictive accuracy}.

\begin{table}[t]
\begin{center}
\begin{tabular}{lrrrrr}
  \hline
 RMSE&  reality & discrepancy &  measurement bias  \\
  \hline
  GaSP           &{0.238}&0.194 &{0.247}   \\
  S-GaSP           &{0.204}& 0.159&{0.214}   \\
  GaSP Stack           &{0.501}&0.483&{0.504}   \\
  S-GaSP Stack         &0.502& 0.484&  {0.504} \\ 
  \hline
\end{tabular}
\end{center}
   \caption{The RMSE of the reality, discrepancy and measurement bias, where the smaller number indicates a smaller error. For  measurement bias, we average the RMSE for each source of data.  }

   \label{tab:prediction_multicalibration}
\end{table}

Finally,  the mean of posterior samples $\theta$ from the GaSP, S-GaSP, GaSP Stack and S-GaSP Stack calibration for calibration parameter $\theta$ are 2.46, 2.73, 2.15, and 2.17 respectively, whereas the truth $\theta=\pi\approx 3.14$. This is a challenging scenario, as the large measurement bias makes estimation hard. We found model calibration by modeling individual sources of data is more accurate than modeling the aggregated data, whereas modeling the aggregated data is more computationally scalable.

\section{Concluding remarks}
We have introduced the  \CRANpkg{RobustCalibration} package for Bayesian model calibration and data inversion. This package has implemented a range of  estimation methods and models, such as posterior sampling and MLE, for no-discrepancy calibration, GaSP and S-GaSP models of the discrepancy function. We implement statistical emulators for approximating computationally expensive computer models with both scalar-valued output or vector-valued output. The package is applicable to observations from a single source, with or without replications, and to observations from multiple sources or having different types. We illustrate our approaches using mathematical models with closed-form expressions or written as differential equations. 

Even though we tried our best to consider different possible scenarios,  a comprehensive statistical package for model calibration and data inversion is an ambitious goal. The \CRANpkg{RobustCalibration} package provides tools for researchers to perform Bayesian model inversion without the need to write emulators or MCMC samplers themselves. We plan to improve the \CRANpkg{RobustCalibration} package with several specific directions in mind. First, we will make the package more computationally scalable for structured data, such as imaging and time series observations. 
Furthermore,  we plan allow users to specify other prior distributions of the parameters and proposal distributions to sample from in future versions of the package. Third, for problems like the Lorzen-96 system, it may be important to emulate the time evolution operator, if forecasting future states is the goal.

\section{Appendix}
\subsection{Auxiliary facts}
\begin{enumerate}[1.]
\item Let $\tilde {\bm R}_{\gamma,\eta}=\mathbf R_{\gamma}/\eta+ \bm \Lambda$ be an $n\times n$ positive definite matrix as a function of parameters $\bm \gamma$ and $\eta$.
 Then for $l=1,...,p_x$: 
\begin{align*}
\frac{\partial \log |\tilde {\bm R}_{\gamma,\eta} | }{\partial  \gamma_l}=\tr\left( \frac{\tilde {\bm R}_{\gamma,\eta}^{-1}}{\eta} \frac{\partial \mathbf R_\gamma}{\partial \bm \gamma_l} \right) \quad \mbox{ and } \quad \frac{\partial \log |\tilde {\bm R}_{\gamma,\eta} | }{\partial \eta}=-\tr\left( \frac{\tilde {\bm R}_{\gamma,\eta}^{-1} \mathbf R_\gamma}{\eta^2} \right). 
\end{align*}
\label{item:derivative_1}
\item  Let $\mathbf H$ be an $n\times q$ full rank matrix with $q<n$, $\mathbf Q=\left(\tilde {\bm R}^{-1}_{\gamma,\eta}- \tilde {\bm R}^{-1}_{\gamma,\eta} \mathbf H^T(\mathbf H^T \tilde {\bm R}^{-1}_{\gamma,\eta} \mathbf H)^{-1}   \mathbf H^T \tilde {\bm R}^{-1}_{\gamma,\eta} \right)$ with $\tilde {\bm R}_{\gamma,\eta}=\mathbf R_{\gamma}/\eta+ \bm \Lambda$ and $\mathbf y$ is an $n\times 1$ vector. Then for $l=1,...,p_x$:  
\begin{align*}
\frac{\partial \log  (\mathbf y^T \mathbf Q_{\gamma,\eta} \mathbf y)   }{\partial  \gamma_l}=-\frac{ \mathbf y^T \mathbf Q_{\gamma,\eta} \frac{\partial \mathbf R_\gamma}{\partial  \gamma_l} \mathbf Q_{\gamma,\eta} \mathbf y }{\eta \mathbf y^T \mathbf Q_{\gamma,\eta} \mathbf y } \quad \mbox{ and } \quad \frac{\partial \log  (\mathbf y^T \mathbf Q_{\gamma,\eta} \mathbf y)   }{\partial   \eta }=\frac{ \mathbf y^T \mathbf Q_{\gamma,\eta} \mathbf R_\gamma \mathbf Q_{\gamma,\eta} \mathbf y }{ \eta^2 \mathbf y^T \mathbf Q_{\gamma,\eta} \mathbf y }.
\end{align*}
\item Suppose  $v_{\bm \theta}$ is an $n\times 1$ vector as a function of $\bm \theta$ and $\mathbf {\tilde R}$ is an $n\times n$ positive definite matrix. Then for $i=1,...,p_{\theta}$: 
\[ \frac{\partial \log(\mathbf v_{\bm \theta}^T \mathbf R^{-1} \mathbf v_{\bm \theta})}{ \partial \theta_l}= \frac{2 \mathbf v_{ \theta}^T\mathbf {\tilde R}^{-1}   \frac{\partial \mathbf v_{\theta}}{\partial  \theta_l}}{\mathbf v^T_{ \theta} \mathbf {\tilde R}^{-1} \mathbf v_{ \theta}}. \] 
\label{item:derivative_2}
\end{enumerate}

\subsection{Likelihood functions and posterior distributions}

\noindent {\bf Posterior distributions of the no-discrepancy calibration}.  
Using the objective prior for the mean and variance parameters, $\pi(\bm \theta_m, \sigma^2_0)\propto 1/\sigma^2_0$,  the full conditional posterior distributions of the trend and variance parameters follow:  
\begin{align*}
 \sigma^{-2}_0 \mid \bm \theta_m,  \bm \theta &\sim \mbox{Gamma}\left(\frac{n}{2},  \frac{(\mathbf y^F-\mathbf f^M_{\theta}- \mathbf H \bm \theta_m )^T\bm \Lambda^{-1}(\mathbf y^F-\mathbf f^M_{\theta}- \mathbf H \bm \theta_m )}{2}  \right), \\ 
\bm \theta_m \mid \sigma^2_0, \bm \theta &\sim \mathcal N\left(   \bm { \hat \theta}^{LS}_m, \,  \sigma^{-2}_0(\mathbf H^T\bm \Lambda^{-1} \mathbf H)^{-1} \right).  
\end{align*}
The posterior distribution of the calibration parameters follows:  
\begin{align*}
p(\bm \theta \mid \bm \theta_m, \bm \sigma^{2}_0, \bm \beta, \eta )\propto \exp\left( - \frac{(\mathbf y^F-\mathbf f^M_{\theta}- \mathbf H \bm \theta_m )^T\bm \Lambda^{-1}(\mathbf y^F-\mathbf f^M_{\theta}- \mathbf H \bm \theta_m )  }{2\sigma^2_0}\right)\pi(\bm \theta),
\end{align*}
where $\pi(\bm \theta)$ is the prior of calibration parameters, where the default prior distribution is the uniform distribution $\pi(\bm \theta)\propto 1$.  We use the Metropolis algorithm to sample the calibration parameters as a block, as we only need to compute the likelihood once in each iteration. The proposal distribution of each coordinate of the calibration parameter vector is chosen as a normal distribution with a pre-specified standard deviation,   proportional to the range of the parameters. The default proportion is $0.05$, and users can adjust the proportion by changing the first $p_{\theta}$ value in the argument \code{sd\_proposal} in \code{rcalibration} function.  

After obtaining $S$ posterior samples with the first $S_0$ burn-in samples, the predictive mean of reality based on calibrated computer model and trend by the no-discrepancy calibration can be computed by:  
\[ \hat \mu(\mathbf x^*)=\frac{1}{S-S_0}\sum^S_{i=S_0+1} f^M(\mathbf x^*,  {\bm \theta}^{(i)} )+\mathbf h(\mathbf x^*)  {\bm \theta_m^{(i)}}.\]
The posterior credible interval of the reality $y^R(\mathbf x^*)$ in the no-discrepancy calibration can be obtained by the empirical quantile of posterior samples $ f^M(\mathbf x^*,  {\bm \theta}^{(i)} )+\mathbf h(\mathbf x^*)  {\bm \theta_m^{(i)}}$. For example, the posterior credible interval of $1-\alpha$ percentile of the data in no-discrepancy calibration can be computed by based on  
$\frac{1}{S-S_0}\sum^S_{i=S_0+1}\left(f^M(\mathbf x^*,  {\bm \theta}^{(i)} )+\mathbf h(\mathbf x^*)  {\bm \theta_m^{(i)}}+\sqrt{\frac{\sigma^{(i)}_0}{ w^*} }Z_{\alpha}\right)$, 
where $Z_{\alpha}$ is an upper $\alpha$ quantile of the standard normal distribution and $w^*$ is the relative test output weight, assumed to be 1 by default. 

\noindent {\bf  Likelihood functions and derivatives in GaSP calibration}. 
The profile likelihood function in GaSP calibration   follows 
\begin{equation}
\mathcal L_K( \bm \theta, \bm \gamma, \eta \mid { \hat {\bm \theta}_m}, \hat \sigma^2_0 ) \propto  |{\mathbf {\tilde R} }|^{-\frac{1}{2}} (S^2_{K})^{-\frac{n}{2}}, 
\label{equ:PL_GP}
\end{equation}
 where $S^2_K=  (\mathbf{y}-\mathbf f^M_{\theta}- \mathbf H^T \hat{\bm \theta}_m  )^T  {{\mathbf { \tilde R}}}^{-1}(\mathbf{y}-\mathbf f^M_{\theta}- \mathbf H^T \hat{\bm \theta}_m  )$.

Denote $\mathbf v_{\bm \theta_m}=(\mathbf{y}-\mathbf f^M_{\theta}- \mathbf H^T \hat{\bm \theta}_m  )$.  Similar to the no-discrepancy calibration, we call the \code{lbfgs} function in the \CRANpkg{nloptr} package (\cite{nloptr2014}) for optimization. 
By facts \ref{item:derivative_1} and \ref{item:derivative_2}, directly differentiating the profile likelihood  function with respect to kernel and nugget parameters in (\ref{equ:PL_GP}) gives:  
\begin{align*}
\frac{\partial \log( \mathcal L_K( \bm \theta, \bm \gamma, \eta \mid {\bm \hat \theta_m}, \hat \sigma^2 ) )}{ \partial \bm \gamma_l } &= -\frac{1}{2}\tr\left( \frac{\tilde {\bm R}_{\gamma,\eta}^{-1}}{\eta} \frac{\partial \mathbf R_\gamma}{\partial \bm \gamma_l} \right)  +\frac{n}{2}\times \frac{ \mathbf v_{\bm \theta_m}^T   \mathbf Q_{\gamma,\eta} \frac{\partial \mathbf R_\gamma}{\partial  \gamma_l} \mathbf Q_{\gamma,\eta} \mathbf v_{\bm \theta_m} }{\eta \mathbf v_{\bm \theta_m}^T \mathbf Q_{\gamma,\eta} \mathbf v_{\bm \theta_m} } \\
\frac{\partial \log( \mathcal L_K( \bm \theta, \bm \gamma, \eta \mid {\bm \hat \theta_m}, \hat \sigma^2 ) )}{ \partial \eta } &= \frac{1}{2}\tr\left( \frac{\tilde {\bm R}_{\gamma,\eta}^{-1} \mathbf R_\gamma}{\eta^2} \right) +\frac{n}{2}\times \frac{ \mathbf v_{\bm \theta_m}^T   \mathbf Q_{\gamma,\eta} \mathbf R_{\gamma}  \mathbf Q_{\gamma,\eta} \mathbf v_{\bm \theta_m} }{\eta^2 \mathbf v_{\bm \theta_m}^T \mathbf Q_{\gamma,\eta} \mathbf v_{\bm \theta_m} } \\
\frac{\partial \log( \mathcal L_K( \bm \theta, \bm \gamma, \eta \mid {\bm \hat \theta_m}, \hat \sigma^2 ) )}{ \partial \theta_l }  &= \frac{n}{2}\times \frac{  \mathbf v_{\bm \theta_m}^T  \mathbf Q_{\gamma,\eta} \frac{\partial \mathbf f^M_\theta}{\partial \theta_l}}{ \mathbf v_{\bm \theta_m}^T   \mathbf Q_{\gamma,\eta}  \mathbf v_{\bm \theta_m} } 
\label{equ:PL}
\end{align*}
for $l=1,...,p_x$ and $i=1,...,p_{\theta}$. When the mean function is zero, one can simply replace $\mathbf Q_{\gamma,\eta}$ by $\tilde {\bm R}_{\gamma,\eta}$ in the formula above to obtain the derivatives.  Besides, when $\frac{\partial \mathbf f^M_\theta}{\partial \theta_l}$ is not available, we use the numerical derivatives to approximate. 

\noindent {\bf Posterior distributions in GaSP calibration}. After marginalizing out the discrepancy function, the marginal distribution follows $\mathbf y^F \sim \mathcal{MN}(\mathbf f^M_{\theta}+\mathbf H \bm \theta_{m},\, \sigma^2_0 \tilde {\mathbf R})$, where $\tilde {\mathbf R}=\mathbf R/\eta+ \bm \Lambda$. We assume an objective prior for the mean and noise variance parameters $\pi(\bm \theta_m, \sigma^2_0)\propto 1/\sigma^2_0$.    The full conditional posterior distributions of the trend and variance parameters follow 
\begin{align*}
\bm \sigma^{-2}_0 \mid \bm \theta_m,  \bm \theta &\sim \mbox{Gamma}\left(\frac{n}{2},  \frac{(\mathbf y^F-\mathbf f^M_{\theta}- \mathbf H \bm \theta_m )^T\bm {\tilde R}^{-1}(\mathbf y^F-\mathbf f^M_{\theta}- \mathbf H \bm \theta_m )}{2}  \right), \\ 
\bm \theta_m \mid \sigma^2_0, \bm \theta &\sim  \mathcal{MN}\left(   \bm { \hat \theta}_m, \,  \sigma^{-2}_0(\mathbf H^T\bm {\tilde R}^{-1} \mathbf H)^{-1} \right), 
\end{align*}
where $\hat {\bm \theta}_m= \left( \mathbf H^T {{\mathbf { \tilde R}}}^{-1}  \mathbf H \right)^{-1} \mathbf H^T {{\mathbf { \tilde R}}}^{-1}{(\mathbf{y}-\mathbf f^M_{\theta})}$. A Gibbs sampler is used to sample $\bm \sigma^{-2}_0$ and $\bm \theta_m$ from the full conditional distribution. 

Given the trend and variance parameters, the posterior distribution of the calibration parameters follows
\begin{align*}
p(\bm \theta \mid \bm \theta_m, \bm \sigma^{2} )\propto \exp\left( - \frac{(\mathbf y^F-\mathbf f^M_{\theta}- \mathbf H \bm \theta_m )^T \mathbf { \tilde R}^{-1}(\mathbf y^F-\mathbf f^M_{\theta}- \mathbf H \bm \theta_m )  }{2\sigma^2_0}\right)\pi(\bm \theta).
\end{align*}
We implement the Metropolis algorithm where the proposal distribution  is  a normal distribution with the mean centered around the current value of the posterior sample. Users can adjust the standard deviation of the proposal distribution by argument \code{sd\_proposal}  in the \code{rcalibration} function.  

Denote the inverse range parameter $\beta_l=1/\gamma_l$, $l=1,...,p_x$. The conditional distribution of the inverse range and nugget parameters follow 
\begin{align*}
p(\bm \beta, \eta \mid  \bm \theta, \bm \theta_m, \bm \sigma^{2}_0 )\propto |\mathbf { \tilde R}|^{-\frac{n}{2}} \exp\left( - \frac{(\mathbf y^F-\mathbf f^M_{\theta}- \mathbf H \bm \theta_m )^T \mathbf { \tilde R}^{-1}(\mathbf y^F-\mathbf f^M_{\theta}- \mathbf H \bm \theta_m )  }{2\sigma^2_0}\right)\pi(\bm \gamma, \eta).
\end{align*}
We define the inverse range parameter $\beta_l=1/\gamma_l$, and assume that the prior for the inverse range and nugget parameters follows the jointly robust (JR) prior \citep{gu2019jointly}
\[\pi(\bm \beta, \eta)=C\left( \sum^{p_x}_{l=1} C_l \beta_l+\eta \right)^a \exp\left\{-b \left( \sum^{p_x}_l C_l \beta_l+\eta \right) \right\}, \]
where $C$ is a normalizing constant. By default, we use the following prior parameters $a=1/2-p_x$, $b=1$, and  $C_l= n^{-1/{p_x}}|x^{max}_{l}-x^{min}_{l}| $, with $x^{max}_{l}$ and $x^{min}_{l}$ being the maximum and minimum values of the observable input at the $l$th coordinate, respectively.  
The default choices of prior parameters induces a small penalty on very large correlation, preventing the identifiability issues of the calibration parameters due to large correlation  \citep{gu2019jointly}. Users can adjust $a$ and $b$ in the \code{rcalibration} function.

We  use the Metropolis algorithm to sample the logarithm of the inverse range parameter $\log( \beta_l)$ and nugget parameter $\log(\eta)$. 
The proposal distribution of each parameter is a normal distribution  centered around the current value  with the standard deviation chosen to be $0.25$ by default. User can change the $p_{\theta}+1$ to  $p_x+p_{\theta}+1$ coordinates of the vector in the argument \code{sd\_proposal}  in the \code{rcalibration} function to specify the standard deviation in the proposal distribution for these parameters.

\noindent {\bf Likelihood function and posterior distributions S-GaSP calibration}. The likelihood function and posterior distribution of  S-GaSP follow from the those in GaSP calibration, by simply replacing the kernel $K$ to $K_{Z_d}$ in Table \ref{tab:kernel}.

\section*{Acknowledgements}  
This research was supported by the National Science Foundation under Award Number DMS-2053423. We thank Xubo Liu for implementing the numerical method for simulating the Lorenz 96 model.

\bibliography{References_2020}

\begin{thebibliography}{40}
\providecommand{\natexlab}[1]{#1}
\providecommand{\url}[1]{\texttt{#1}}
\expandafter\ifx\csname urlstyle\endcsname\relax
  \providecommand{\doi}[1]{doi: #1}\else
  \providecommand{\doi}{doi: \begingroup \urlstyle{rm}\Url}\fi

\bibitem[Agram and Simons(2015)]{agram2015noise}
P.~Agram and M.~Simons.
\newblock A noise model for {I}n{SAR} time series.
\newblock \emph{Journal of Geophysical Research: Solid Earth}, 120\penalty0
  (4):\penalty0 2752--2771, 2015.

\bibitem[Anderson et~al.(2019)Anderson, Johanson, Patrick, Gu, Segall, Poland,
  Montgomery-Brown, and Miklius]{anderson2019magma}
K.~R. Anderson, I.~A. Johanson, M.~R. Patrick, M.~Gu, P.~Segall, M.~P. Poland,
  E.~K. Montgomery-Brown, and A.~Miklius.
\newblock Magma reservoir failure and the onset of caldera collapse at
  {K}{\=\i}lauea volcano in 2018.
\newblock \emph{Science}, 366\penalty0 (6470), 2019.

\bibitem[Arendt et~al.(2012)Arendt, Apley, and Chen]{arendt2012quantification}
P.~D. Arendt, D.~W. Apley, and W.~Chen.
\newblock Quantification of model uncertainty: calibration, model discrepancy,
  and identifiability.
\newblock \emph{Journal of Mechanical Design}, 134\penalty0 (10):\penalty0
  100908, 2012.

\bibitem[Bayarri et~al.(2007{\natexlab{a}})Bayarri, Berger, Cafeo,
  Garcia-Donato, Liu, Palomo, Parthasarathy, Paulo, Sacks, and
  Walsh]{bayarri2007computer}
M.~Bayarri, J.~Berger, J.~Cafeo, G.~Garcia-Donato, F.~Liu, J.~Palomo,
  R.~Parthasarathy, R.~Paulo, J.~Sacks, and D.~Walsh.
\newblock Computer model validation with functional output.
\newblock \emph{The Annals of Statistics}, 35\penalty0 (5):\penalty0
  1874--1906, 2007{\natexlab{a}}.

\bibitem[Bayarri et~al.(2007{\natexlab{b}})Bayarri, Berger, Paulo, Sacks,
  Cafeo, Cavendish, Lin, and Tu]{bayarri2007framework}
M.~J. Bayarri, J.~O. Berger, R.~Paulo, J.~Sacks, J.~A. Cafeo, J.~Cavendish,
  C.-H. Lin, and J.~Tu.
\newblock A framework for validation of computer models.
\newblock \emph{Technometrics}, 49\penalty0 (2):\penalty0 138--154,
  2007{\natexlab{b}}.

\bibitem[Box and Coutie(1956)]{box1956application}
G.~Box and G.~Coutie.
\newblock Application of digital computers in the exploration of functional
  relationships.
\newblock \emph{Proceedings of the IEE-Part B: Radio and Electronic
  Engineering}, 103\penalty0 (1S):\penalty0 100--107, 1956.

\bibitem[Brajard et~al.(2020)Brajard, Carrassi, Bocquet, and
  Bertino]{brajard2020combining}
J.~Brajard, A.~Carrassi, M.~Bocquet, and L.~Bertino.
\newblock Combining data assimilation and machine learning to emulate a
  dynamical model from sparse and noisy observations: A case study with the
  {L}orenz 96 model.
\newblock \emph{Journal of Computational Science}, 44:\penalty0 101171, 2020.

\bibitem[Carmassi et~al.(2018)Carmassi, Barbillon, Chiodetti, Keller, and
  Parent]{carmassi2018calico}
M.~Carmassi, P.~Barbillon, M.~Chiodetti, M.~Keller, and E.~Parent.
\newblock Cali{C}o: a {R} package for {B}ayesian calibration.
\newblock \emph{arXiv preprint arXiv:1808.01932}, 2018.

\bibitem[Chang et~al.(2016)Chang, Haran, Applegate, and
  Pollard]{chang2016calibrating}
W.~Chang, M.~Haran, P.~Applegate, and D.~Pollard.
\newblock Calibrating an ice sheet model using high-dimensional binary spatial
  data.
\newblock \emph{Journal of the American Statistical Association}, 111\penalty0
  (513):\penalty0 57--72, 2016.

\bibitem[Chang et~al.(2022)Chang, Konomi, Karagiannis, Guan, and
  Haran]{chang2022ice}
W.~Chang, B.~A. Konomi, G.~Karagiannis, Y.~Guan, and M.~Haran.
\newblock Ice model calibration using semicontinuous spatial data.
\newblock \emph{The Annals of Applied Statistics}, 16\penalty0 (3):\penalty0
  1937--1961, 2022.

\bibitem[Fang et~al.(2022)Fang, Gu, and Wu]{fang2022reliable}
X.~Fang, M.~Gu, and J.~Wu.
\newblock Reliable emulation of complex functionals by active learning with
  error control.
\newblock \emph{The Journal of Chemical Physics}, 157\penalty0 (21):\penalty0
  214109, 2022.

\bibitem[Gu(2019)]{gu2019jointly}
M.~Gu.
\newblock Jointly robust prior for {G}aussian stochastic process in emulation,
  calibration and variable selection.
\newblock \emph{Bayesian Analysis}, 14\penalty0 (1), 2019.

\bibitem[Gu and Wang(2018)]{gu2018sgasp}
M.~Gu and L.~Wang.
\newblock Scaled {G}aussian stochastic process for computer model calibration
  and prediction.
\newblock \emph{SIAM/ASA Journal on Uncertainty Quantification}, 6\penalty0
  (4):\penalty0 1555--1583, 2018.

\bibitem[Gu et~al.(2019)Gu, Palomo, and Berger]{gu2018robustgasp}
M.~Gu, J.~Palomo, and J.~O. Berger.
\newblock {RobustGaSP: Robust Gaussian Stochastic Process Emulation in R}.
\newblock \emph{{The R Journal}}, 11\penalty0 (1):\penalty0 112--136, 2019.
\newblock \doi{10.32614/RJ-2019-011}.

\bibitem[Gu et~al.(2022)Gu, Xie, and Wang]{gu2022theoretical}
M.~Gu, F.~Xie, and L.~Wang.
\newblock A theoretical framework of the scaled {G}aussian stochastic process
  in prediction and calibration.
\newblock \emph{SIAM/ASA Journal on Uncertainty Quantification}, 10\penalty0
  (4):\penalty0 1435--1460, 2022.

\bibitem[Gu et~al.(2023)Gu, Anderson, and McPhillips]{gu2022calibration}
M.~Gu, K.~Anderson, and E.~McPhillips.
\newblock Calibration of imperfect geophysical models by multiple satellite
  interferograms with measurement bias.
\newblock \emph{Technometrics, In Press}, 2023.
\newblock \doi{10.1080/00401706.2023.2182365}.

\bibitem[Hankin(2005)]{hankin2005introducing}
R.~K. Hankin.
\newblock Introducing {BACCO}, an {R} bundle for {B}ayesian analysis of
  computer code output.
\newblock \emph{Journal of Statistical Software}, 14:\penalty0 1--21, 2005.

\bibitem[Higdon et~al.(2008)Higdon, Gattiker, Williams, and
  Rightley]{higdon2008computer}
D.~Higdon, J.~Gattiker, B.~Williams, and M.~Rightley.
\newblock Computer model calibration using high-dimensional output.
\newblock \emph{Journal of the American Statistical Association}, 103\penalty0
  (482):\penalty0 570--583, 2008.

\bibitem[Kennedy and O'Hagan(2001)]{kennedy2001bayesian}
M.~C. Kennedy and A.~O'Hagan.
\newblock Bayesian calibration of computer models.
\newblock \emph{Journal of the Royal Statistical Society: Series B (Statistical
  Methodology)}, 63\penalty0 (3):\penalty0 425--464, 2001.

\bibitem[Li et~al.(2022)Li, Zhou, Sebastian, Wu, and Gu]{li2022efficient}
H.~Li, M.~Zhou, J.~Sebastian, J.~Wu, and M.~Gu.
\newblock Efficient force field and energy emulation through partition of
  permutationally equivalent atoms.
\newblock \emph{The Journal of Chemical Physics}, 156\penalty0 (18):\penalty0
  184304, 2022.

\bibitem[Liu and Nocedal(1989)]{liu1989limited}
D.~C. Liu and J.~Nocedal.
\newblock On the limited memory {BFGS} method for large scale optimization.
\newblock \emph{Mathematical programming}, 45\penalty0 (1-3):\penalty0
  503--528, 1989.

\bibitem[Liu et~al.(2009)Liu, Bayarri, and Berger]{liu2009modularization}
F.~Liu, M.~Bayarri, and J.~Berger.
\newblock Modularization in {B}ayesian analysis, with emphasis on analysis of
  computer models.
\newblock \emph{Bayesian Analysis}, 4\penalty0 (1):\penalty0 119--150, 2009.

\bibitem[Lorenz(1996)]{lorenz1996predictability}
E.~N. Lorenz.
\newblock Predictability: A problem partly solved.
\newblock In \emph{Proc. Seminar on predictability}, volume~1, 1996.

\bibitem[Ma et~al.(2022)Ma, Karagiannis, Konomi, Asher, Toro, and
  Cox]{ma2022multifidelity}
P.~Ma, G.~Karagiannis, B.~A. Konomi, T.~G. Asher, G.~R. Toro, and A.~T. Cox.
\newblock Multifidelity computer model emulation with high-dimensional output:
  An application to storm surge.
\newblock \emph{Journal of the Royal Statistical Society Series C: Applied
  Statistics}, 71\penalty0 (4):\penalty0 861--883, 2022.

\bibitem[MacDonald et~al.(2015)MacDonald, Ranjan, Chipman,
  et~al.]{macdonald2015gpfit}
B.~MacDonald, P.~Ranjan, H.~Chipman, et~al.
\newblock Gpfit: An {R} package for fitting a {G}aussian process model to
  deterministic simulator outputs.
\newblock \emph{Journal of Statistical Software}, 64\penalty0 (i12), 2015.

\bibitem[Maclean and Spiller(2020)]{maclean2020surrogate}
J.~Maclean and E.~T. Spiller.
\newblock A surrogate-based approach to nonlinear, non-{G}aussian joint
  state-parameter data assimilation.
\newblock \emph{arXiv preprint arXiv:2012.04793}, 2020.

\bibitem[Palomo et~al.(2015)Palomo, Paulo, Garc{\'\i}a-Donato,
  et~al.]{palomo2015save}
J.~Palomo, R.~Paulo, G.~Garc{\'\i}a-Donato, et~al.
\newblock Save: an {R} package for the statistical analysis of computer models.
\newblock \emph{Journal of Statistical Software}, 64\penalty0 (13):\penalty0
  1--23, 2015.

\bibitem[Paulo et~al.(2012)Paulo, Garc{\'\i}a-Donato, and
  Palomo]{paulo2012calibration}
R.~Paulo, G.~Garc{\'\i}a-Donato, and J.~Palomo.
\newblock Calibration of computer models with multivariate output.
\newblock \emph{Computational Statistics and Data Analysis}, 56\penalty0
  (12):\penalty0 3959--3974, 2012.

\bibitem[Plumlee(2017)]{plumlee2017bayesian}
M.~Plumlee.
\newblock Bayesian calibration of inexact computer models.
\newblock \emph{Journal of the American Statistical Association}, 112\penalty0
  (519):\penalty0 1274--1285, 2017.

\bibitem[Roustant et~al.(2012)Roustant, Ginsbourger, and
  Deville]{roustant2012dicekriging}
O.~Roustant, D.~Ginsbourger, and Y.~Deville.
\newblock Dicekriging, {D}iceoptim: Two {R} packages for the analysis of
  computer experiments by kriging-based metamodeling and optimization.
\newblock \emph{Journal of statistical software}, 51:\penalty0 1--55, 2012.

\bibitem[Sacks et~al.(1989)Sacks, Welch, Mitchell, Wynn,
  et~al.]{sacks1989design}
J.~Sacks, W.~J. Welch, T.~J. Mitchell, H.~P. Wynn, et~al.
\newblock Design and analysis of computer experiments.
\newblock \emph{Statistical science}, 4\penalty0 (4):\penalty0 409--423, 1989.

\bibitem[Santner et~al.(2003)Santner, Williams, and Notz]{santner2003design}
T.~J. Santner, B.~J. Williams, and W.~I. Notz.
\newblock \emph{The design and analysis of computer experiments}.
\newblock Springer Science \& Business Media, 2003.

\bibitem[Simakov et~al.(2019)Simakov, Jones-Ivey, Akhavan-Safaei, Aghakhani,
  Jones, and Patra]{simakov2019modernizing}
N.~A. Simakov, R.~L. Jones-Ivey, A.~Akhavan-Safaei, H.~Aghakhani, M.~D. Jones,
  and A.~K. Patra.
\newblock Modernizing {T}itan2{D}, a parallel {AMR} geophysical flow code to
  support multiple rheologies and extendability.
\newblock In \emph{International Conference on High Performance Computing},
  pages 101--112. Springer, 2019.

\bibitem[Soetaert et~al.(2010)Soetaert, Petzoldt, and
  Setzer]{soetaert2010solving}
K.~Soetaert, T.~Petzoldt, and R.~W. Setzer.
\newblock Solving differential equations in {R}: package de{S}olve.
\newblock \emph{Journal of statistical software}, 33:\penalty0 1--25, 2010.

\bibitem[Tuo and Wu(2015)]{tuo2015efficient}
R.~Tuo and C.~J. Wu.
\newblock Efficient calibration for imperfect computer models.
\newblock \emph{The Annals of Statistics}, 43\penalty0 (6):\penalty0
  2331--2352, 2015.

\bibitem[Wickham(2011)]{wickham2011ggplot2}
H.~Wickham.
\newblock ggplot2.
\newblock \emph{Wiley Interdisciplinary Reviews: Computational Statistics},
  3\penalty0 (2):\penalty0 180--185, 2011.

\bibitem[Williamson et~al.(2013)Williamson, Goldstein, Allison, Blaker,
  Challenor, Jackson, and Yamazaki]{williamson2013history}
D.~Williamson, M.~Goldstein, L.~Allison, A.~Blaker, P.~Challenor, L.~Jackson,
  and K.~Yamazaki.
\newblock History matching for exploring and reducing climate model parameter
  space using observations and a large perturbed physics ensemble.
\newblock \emph{Climate dynamics}, 41\penalty0 (7-8):\penalty0 1703--1729,
  2013.

\bibitem[Wong et~al.(2017)Wong, Storlie, and Lee]{wong2017frequentist}
R.~K. Wong, C.~B. Storlie, and T.~Lee.
\newblock A frequentist approach to computer model calibration.
\newblock \emph{Journal of the Royal Statistical Society: Series B (Statistical
  Methodology)}, 79:\penalty0 635--648, 2017.

\bibitem[Ypma(2014)]{nloptr2014}
J.~Ypma.
\newblock \emph{nloptr: {R} interface to NLopt}, 2014.
\newblock URL \url{https://CRAN.R-project.org/package=nloptr}.
\newblock R package version 1.0.4.

\bibitem[Zebker et~al.(1997)Zebker, Rosen, and Hensley]{Zebker1997}
H.~A. Zebker, P.~A. Rosen, and S.~Hensley.
\newblock {Atmospheric effects in interferometric synthetic aperture radar
  surface deformation and topographic maps}.
\newblock \emph{Journal of Geophysical Research: Solid Earth}, 102\penalty0
  (B4):\penalty0 7547--7563, apr 1997.
\newblock ISSN 01480227.
\newblock \doi{10.1029/96JB03804}.
\newblock URL \url{http://doi.wiley.com/10.1029/96JB03804}.

\end{thebibliography}

\address{
  Mengyang Gu\\
    University of California, Santa Barbara\\
  Department of Statistics and Applied Probability\\
  Santa Barbara, California, USA\\
   \email{mengyang@pstat.ucsb.edu}}

\end{article}

\end{document}